\documentclass[12pt]{iopart}
\usepackage{iopams}  

\usepackage{graphicx}
\usepackage{setstack}
\usepackage{appendix}
\usepackage{enumerate}

\newcommand{\ket}[1]{|#1\rangle}
\newcommand{\bra}[1]{\langle#1|}

\begin{document}

\newtheorem{theorem}{Theorem}
\newtheorem{lemma}[theorem]{Lemma}
\newtheorem{corollary}[theorem]{Corollary}
\newtheorem{proposition}[theorem]{Proposition}
\newtheorem{definition}[theorem]{Definition}
\newtheorem{example}[theorem]{Example}
\newtheorem{conjecture}[theorem]{Conjecture}

\title{Driving two atoms in an optical cavity into an entangled steady state using engineered decay}
\date{\today}
\author{Florentin Reiter, Michael J Kastoryano and Anders S S\o rensen}
\address{QUANTOP, Danish Quantum Optics Center, Niels Bohr Institute, Blegdamsvej 17, DK-2100 Copenhagen \O, Denmark}
\ead{reiter@nbi.dk}
\begin{abstract}
We propose various schemes for dissipative preparation of a maximally entangled steady state of two atoms in an optical cavity. Harnessing the natural decay processes of cavity photon loss and spontaneous emission, we use an effective operator formalism to identify and engineer effective decay processes, which reach an entangled steady state of two atoms as the unique fixed point of the dissipative time evolution. We investigate various aspects that are crucial for the experimental implementation of our schemes in present-day and future cavity quantum electrodynamics systems and analytically derive the optimal parameters, the error scaling and the speed of convergence of our protocols. Our study shows promising performance of our schemes for existing cavity experiments and favorable scaling of fidelity and speed with respect to the cavity parameters.\\
\end{abstract}
\vspace{-1.0cm}

\vspace{-0.8cm}
\renewcommand*\contentsname{}
\begin{tableofcontents}
\end{tableofcontents}
\vspace{-10cm}
\title[Driving two atoms in a cavity into an entangled steady state using engineered decay]{}

\clearpage

\section{Introduction}
The reliable and efficient preparation of entangled states has been one of the main tasks in quantum information science since the birth of the field. The effort has been driven on the one hand by the desire to understand these quintessentially non-classical states of matter, and on the other by their promise as building blocks for quantum information processing tasks. In particular, bipartite maximally entangled states constitute the standard of entanglement theory, which in turn is believed to be the main ingredient responsible for the additional information processing power of quantum machines over classical ones.\\
Since the advent of quantum information science, noise has been considered a detrimental element in a physical setup, causing decoherence which must at all cost be avoided. A few years ago, however, it has been suggested that dissipative noise can be used as a resource for quantum information processing, abetting in the preparation of entangled states.  For instance, Refs. \cite{Kim, SchneiderMilburn, Basharov, Jakobczyk, Braun, Benatti1, Benatti2} showed that the dissipative dynamics of two atoms coupled to a common reservoir could lead to entanglement. These initial ideas were generalized in Refs. \cite{VWC, Diehl, Kraus} to show that indeed a very general class of states and quantum informations tasks could be realized by dissipation.
Since then many quantum information processing tasks have been reconsidered within the framework of dissipative state engineering. For instance, universal quantum computation \cite{VWC}, entanglement distillation and quantum repeaters \cite{DissRepeater}, quantum memories \cite{DissMemory}, quantum simulators \cite{Barreiro, DissSimulation} and various forms of entangled state preparation \cite{PCZ, PHBK, Clark, Parkins, VB, WS, CBK, KRS, MPC, Krauter, Busch, AGZ, DGM}, have all been shown possible using only dissipation as a resource. The physical systems which have so-far been proposed or used for dissipative state preparation are: Cavity quantum electrodynamics (QED) \cite{PHBK, Clark, Parkins, WS, KRS, Busch}, ion traps \cite{SchneiderMilburn, Barreiro, DissSimulation, PCZ}, optical lattices \cite{Diehl, VB},  atomic ensembles \cite{Parkins, MPC, Krauter}, and plasmonic waveguides \cite{AGZ, DGM}. The first experimental studies along these lines \cite{Barreiro, Krauter} have shown these new ideas to be realistic and promising as a new path for harnessing the potential of quantum information.\\
In this paper, we consider two $\Lambda$-atoms trapped in a single mode cavity QED setup \cite{DRBH, ZG, PW, SM} coherently driven by a classical optical field and a microwave or Raman field. We demonstrate that a maximally entangled steady state (stationary state) of the two atoms can be prepared dissipatively with a very high fidelity by several qualitatively different state preparation mechanisms. Which mechanism is desirable ultimately depends on the strengths and limitations of the experiment at hand \cite{Kubanek1, Koch, Boozer1, Boozer2, Schuster}. In each of the schemes, the two atoms are rapidly driven into a singlet state, independent of the initial state of the system, and without need for any unitary feedback control. Consequently, the lifetime of the state is dictated by the lifetime of the experiment. As maximally entangled states are an important resource in many quantum information processing protocols (ex: repeaters, cryptography), having access to a reliable source can not be overestimated.\\
Below we will consider a number of schemes:
\begin{enumerate}
\item A scheme that employs spontaneous emission mediated by the dark state of the atom-cavity interaction. This scheme leads to the highest fidelity of the entangled steady state among the presented schemes. In Sec. \ref{SectionS1} we analyze this scheme in full detail and derive benchmarks for error and speed of the protocol in Sec. \ref{SectionSpeed}.
\item In Sec. \ref{SectionT0} we present various schemes suitable for existing cavity QED experiments which do not have transversal confinement of the atoms, cf. Ref. \cite{Kubanek1}.
\item For completeness we discuss the scheme of Ref. \cite{KRS} developed by us previously which uses engineered cavity decay (Sec. \ref{AppS0}), and an adaptation of a dissipative protocol presented in Ref. \cite{WS} to $\Lambda$ atoms in optical cavities (Sec. \ref{AppWS}).
\end{enumerate}
For each of the schemes, we identify the relevant interactions by systematically truncating the Hilbert space of the problem by an effective operator formalism based on second order perturbation theory and adiabatic elimination of the excited states \cite{EO}. This gives us an effective master equation for the schemes (i-iii), from which all of the desired performance measures can be analytically derived. For the scheme (i), we analytically derive the optimal steady-state fidelity and the convergence time as a function of the system parameters in Sec. \ref{SectionSpeed}. In Sec. \ref{AppAsymm} we analyze its robustness against a difference in the coupling of the atoms to the cavity which is present even in state-of-the-art optical cavities. In all cases the analytic results are verified by numerical simulations. The results obtained are collected and compared in Sec. \ref{SectionComparison}.\\
We note that our studies are in a sense related to the more abstract protocols presented in Refs. \cite{Kim, SchneiderMilburn, Basharov, Jakobczyk, Braun, Benatti1, Benatti2}, showing that a set of entangled states of two two-level systems coupled to a common Markovian bath can be reached by purely dissipative means. However, as we aim to demonstrate schemes for concrete cavity QED experiments, we focus on maximally entangled states of three-level systems. Our studies can be seen as a step towards the desired general tools outlined in Refs. \cite{VWC, Diehl, Kraus}. A highly related scheme was also presented in Ref. \cite{Busch}, but a comparison to this scheme is beyond the scope of this work.
\begin{figure*}[ht]
\centering
\includegraphics[width=13cm]{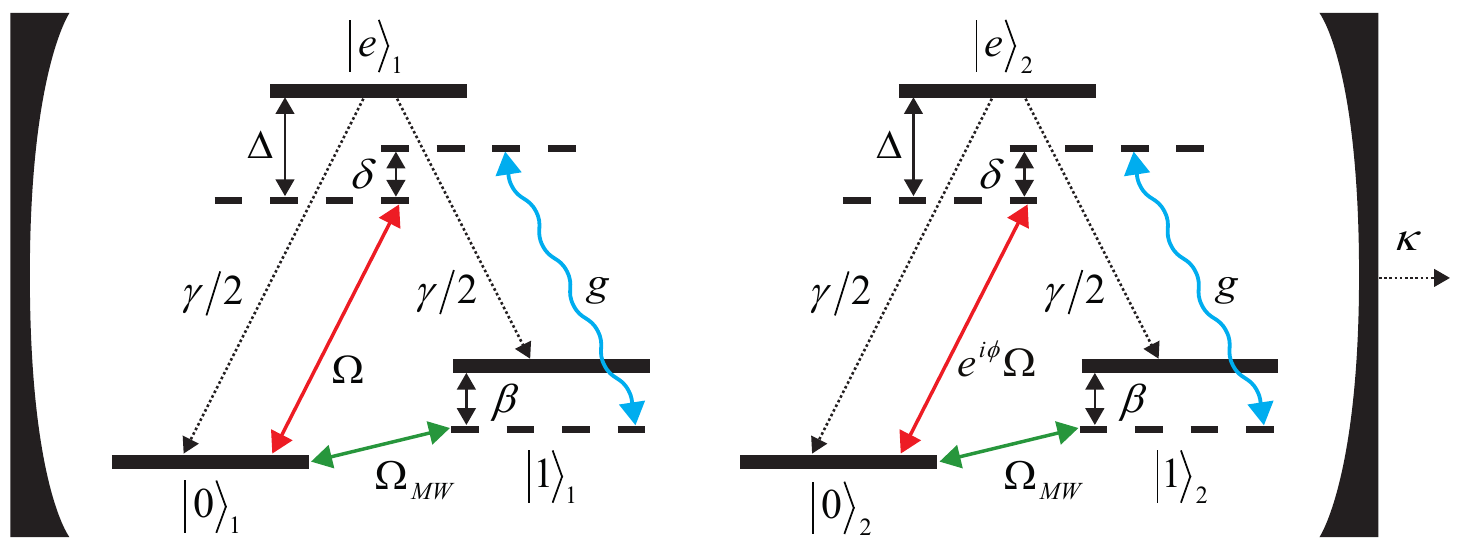}
\caption{Cavity QED setup for dissipative preparation of an entangled state between two $\Lambda$-type atoms in an optical cavity. Coherent driving $\Omega$ is performed from ground state $\ket{0}$ to the excited state $\ket{e}$. Atom-cavity interaction $g$ takes place between $\ket{e}$ and ground state $\ket{1}$; the ground states are coupled by a microwave or Raman transition $\Omega_{\rm MW}$. Spontaneous emission and cavity photon loss are present as sources of decay.}
\label{FigSetup}
\end{figure*}

\section{Cavity QED setups for dissipative preparation of entanglement}
\label{SectionSetup}
\begin{figure*}[t]
\centering
\includegraphics[width=14cm]{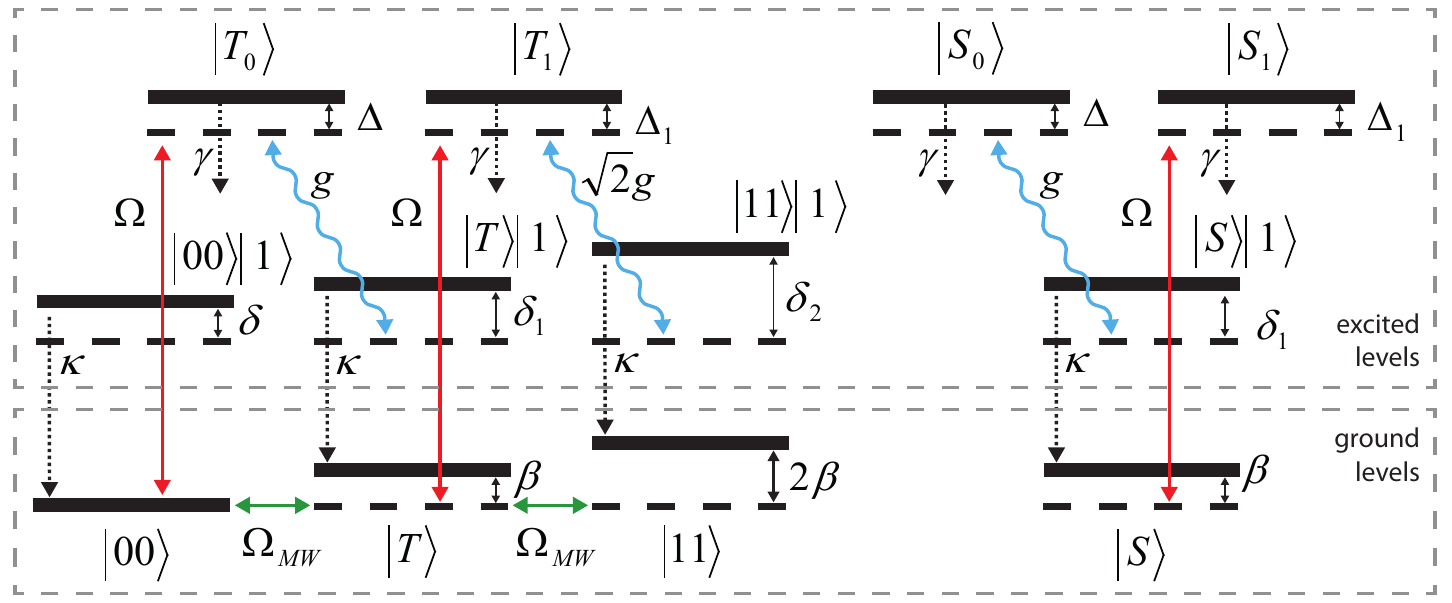}
\caption{Coherent and dissipative interactions between ground and excited states. Ground states are coherently excited by an optical field ($\Omega$) to excited atomic states (straight red arrow). Exchange of excitation via the atom-cavity interaction ($g$) couples these to cavity-excited states (wiggled blue arrow). Ground states are coupled by a microwave or Raman transition $\Omega_{\rm MW}$. Atomic excitations decay by spontaneous emission at a rate of $\gamma$ and cavity loss occurs at a rate of $\kappa$ (both indicated with dotted arrows). The ground to excited subspace interactions are drawn for $\phi=0$ where excitation happens inside the triplet/singlet subspaces, whereas $\phi=\pi$ leads to crossings between the triplet and singlet sectors. The detunings are defined in the text.}
\label{FigInteraction}
\end{figure*}

We consider a single-mode cavity QED system consisting of two distantly trapped $\Lambda$-type atoms with ground states $\ket{0}$ and $\ket{1}$ and an excited state $\ket{e}$. These levels are coupled by a laser, a microwave field or Raman transition, and a cavity mode. Within the dipole and rotating wave approximation the couplings of this system are described by a Hamiltonian
\begin{eqnarray}
\hat{H} = \hat{H}_{\rm 0} + \hat{H}_{\rm ac} + \hat{H}_{\rm laser} + \hat{H}_{\rm MW}.
\end{eqnarray}
We assume that the level splittings are the same for the two atoms, and do not fluctuate in time. This can for instance be achieved by cooling the atoms to the ground states of identical trapping potentials or by using `magic wavelength traps'. Then the Hamiltonian for the atoms and the cavity mode is given by
\begin{eqnarray}
\hat{H}_{0} = \omega_{\rm c} \hat{a}^{\dagger}\hat{a} + \sum_{j=1,2} \left(\omega_0 \ket{0}_j\bra{0} + \omega_1 \ket{1}_j\bra{1} + \omega_{\rm e} \ket{e}_j\bra{e}\right).
\end{eqnarray}
The couplings of the levels are expressed by the interaction Hamiltonians
\begin{eqnarray}
\hat{H}_{\rm laser} = \frac{\Omega}{2}  e^{-i \omega_{\rm laser} t} \left(\ket{e}_1\bra{0}+e^{i\phi} \ket{e}_2\bra{0}\right) + H. c.\\
\hat{H}_{\rm MW} = \frac{\Omega_{\rm MW}}{2} \sum_{j=1,2}  e^{-i \omega_{\rm MW} t} \ket{1}_j\bra{0} + H. c.\\
\hat{H}_{\rm ac} = g \sum_{j=1,2} \hat{a} \ket{e}_j\bra{1} + H. c.
\end{eqnarray}
The coherent laser field $\hat{H}_{\rm laser}$ drives the transition between the ground state $\ket{0}$ and the excited state $\ket{e}$ with resonant Rabi frequency $\Omega$. The angle $\phi$ determines the phase difference of the driving field for the two atoms with respect to the atom-cavity coupling; for convenience we assume it on the driving of the second atom. The two ground states $\ket{0}$ and $\ket{1}$ are coupled by means of a coherent microwave field or Raman transition $\hat{H}_{\rm MW}$ of Rabi frequency $\Omega_{\rm MW}$. The atom-cavity interaction $\hat{H}_{\rm ac}$ describes that the cavity field $\{\hat{a},\hat{a}^\dag\}$ couples the $\ket{1}$ and $\ket{e}$ states with a strength of $g$ and an identical phase for both atoms. Assuming that the ground and excited subspace are coupled perturbatively, the system Hamiltonian can be structured according to
\begin{eqnarray}
\label{EqH1}
\hat{H} = \hat{H}_{\rm g} + \hat{H}_{\rm e} + \hat{V}_+ + \hat{V}_-,
\end{eqnarray}
with $\hat{H}_{\rm g}$ ($\hat{H}_{\rm e}$) as the Hamiltonian of the ground (excited) subspace and $\hat{V}=\hat{V}_+ + \hat{V}_-$ ($\hat{V}_-=\hat{V}^\dagger_+$) as the perturbative (de-)excitation term between the ground and excited subspaces. We change into a frame rotating at the frequency of the level $\ket{0}$, $\omega_0$, and the frequencies of the laser and the microwave, $\omega_{\rm laser}$ and $\omega_{\rm MW}$, 
to obtain the time-independent couplings illustrated in Fig. \ref{FigSetup},
\begin{eqnarray}
\hat{H}_{\rm g} = \frac{\Omega_{\rm MW}}{2} \sum_{j=1,2} \left(\ket{1}_j\bra{0} + H. c.\right) + \beta \sum_{j=1,2} \ket{1}_j\bra{1}\\
\hat{H}_{\rm e} = \Delta \sum_{j=1,2} \ket{e}_j\bra{e} + \delta \hat{a}^{\dagger}\hat{a} + g  \sum_{j=1,2} \left(\hat{a}^\dagger \ket{1}_j\bra{e} + H. c.\right)\\
\hat{V}_+ = \frac{\Omega}{2} \left(\ket{e}_1\bra{0}+e^{i\phi} \ket{e}_2\bra{0}\right), ~ \hat{V}_-=\hat{V}^\dagger_+.
\label{EqH2}
\end{eqnarray}
Here, $\Delta \equiv \omega_{\rm e} - \omega_0 - \omega_{\rm laser}$ and $\beta \equiv \omega_{1} - \omega_{0} - \omega_{\rm MW}$ are the detunings of the laser and of the microwave field, respectively; a cavity excitation has an energy of $\delta \equiv \omega_{\rm c} - \omega_{\rm laser} + \omega_{\rm MW}$.\\
The noise processes resulting from interactions between the system and the environment are assumed Markovian so that the time evolution of the system can be described by a master equation $\dot{\rho}=\mathcal{L}[\rho]$ with a Liouvillian in Lindblad form
\begin{equation}
\label{EqFullMaster}
\dot{\rho}=\mathcal{L}[\rho]=-i\left[\hat{H},\rho\right]+\sum_k \hat{L}_{k} \rho \hat{L}^\dagger_{k} - \frac{1}{2}\left(\hat{L}^\dagger_{k}\hat{L}_{k}\rho+\rho\hat{L}^\dagger_{k}\hat{L}_{k}\right).
\end{equation}
The Lindblad operators $\hat{L}_k$ are associated with the following naturally occurring noise processes: (i) loss of a cavity excitation, $\hat{L}_{\kappa}=\sqrt{\kappa}\hat{a}$ with a photon decay rate of $\kappa$, and (ii) decay by spontaneous emission from the excited atomic state $\ket{e}_j$ into ground states $\ket{0}_j$ and $\ket{1}_j$, with Lindblad operators $\hat{L}_{\gamma,0,j}=\sqrt{\gamma/2}\ket{0}_j\bra{e}$ and $\hat{L}_{\gamma,1,j}=\sqrt{\gamma/2}\ket{1}_j\bra{e}$, respectively. Given that the separation of the atoms will typically be more than one wavelength for typical experimental conditions we neglect collective components of the spontaneous emission. Furthermore, we see from the arguments below that the bandwidth of the laser plays a minor role, as long as it is kept within the linewidth of the transition we want to drive, e. g. $\sim 6$ MHz for the parameters of Ref. \cite{Kubanek1}. Note that for simplicity we assume equal rates $\gamma/2$ into the two ground states; an asymmetric decay of $\ket{e}$ does not modify the results significantly and can even be used to improve the protocols. We ignore the influence from other sources of noise than spontaneous emission and cavity decay, such as dephasing due to stray fields or blackbody radiation. This is justified if the coherence time of the hyperfine transition exceeds the preparation time. As we shall see later the preparation time is on the order of $\mu$s justifying this approximation for most experiments.\\
As a measure for the quality of the cavity we introduce the \textit{cooperativity} parameter, $C = \frac{g^2}{\gamma\kappa}$. As we will see later the cooperativity is the main parameter quantifying how well the entanglement protocols work. We note here, that we define $\gamma$ as the decay of population of the excited state and $\kappa$ as the photon loss rate of the cavity, which differ from polarization and field decay rates which are also commonly referred to as $\left(\gamma,\kappa\right)$ \cite{Kubanek1} by a factor of two each; hence, for the parameters of Ref. \cite{Kubanek1} we get $C \approx 17$. In this work, we will always assume strong coupling $C  \gg 1$, but we distinguish between the regimes of weak driving $\left(\Omega, \Omega_{\rm MW}, \beta\right) \ll \left(g, \kappa, \gamma\right)$ and increased driving $\left(\Omega, \Omega_{\rm MW},\beta\right) \lesssim \left(\kappa, \gamma\right)$.\\
\\
In the following, it will be convenient to work in the triplet-singlet basis of the ground states: $\{\ket{00},\ket{11},\ket{T},\ket{S}\}$, where $\ket{00}=\ket{0}_1\ket{0}_2$, $\ket{11}=\ket{1}_1\ket{1}_2$, $\ket{T}=\frac{1}{\sqrt{2}}\left(\ket{0}_1\ket{1}_2+\ket{1}_1\ket{0}_2\right)$, and $\ket{S}=\frac{1}{\sqrt{2}}\left(\ket{0}_1\ket{1}_2-\ket{1}_1\ket{0}_2\right)$; the latter is the desired maximally entangled singlet state for all the protocols that we investigate. We further define the following excited states which will appear throughout the article: $\ket{T_0}=\frac{1}{\sqrt{2}}\left(\ket{0}_1\ket{e}_2+\ket{e}_1\ket{0}_2\right)$, $\ket{S_0}=\frac{1}{\sqrt{2}}\left(\ket{0}_1\ket{e}_2-\ket{e}_1\ket{0}_2\right)$, $\ket{T_1}=\frac{1}{\sqrt{2}}\left(\ket{1}_1\ket{e}_2+\ket{e}_1\ket{1}_2\right)$ and $\ket{S_1}=\frac{1}{\sqrt{2}}\left(\ket{1}_1\ket{e}_2-\ket{e}_1\ket{1}_2\right)$.
The excited states of the cavity read $\ket{00}\ket{1}$, $\ket{11}\ket{1}$, $\ket{T}\ket{1}$, $\ket{S}\ket{1}$. We will truncate the Hilbert space by excluding states with two or more excitations (of the cavity or of the atoms), as we always work in the perturbative regime with weak driving in our analytical calculations. The states and their couplings (for $\phi=0$) are shown in Fig. \ref{FigInteraction}.

\section{Effective open system dynamics}
\label{SectionMethods}
The key to establishing an entangled ground state by dissipative state preparation is to identify and engineer decay processes present in the physical system in a systematic and reliable way. In the following section we present the methods that are used throughout this article to model the open quantum system at hand. We introduce an effective operator formalism that allows to reduce the unitary and dissipative dynamics of the open-quantum system to the non-decaying ground states and permits us to tailor the effective decay processes to achieve a desired steady state.

\subsection*{Complex energies and non-Hermitian time evolution of the excited states}
In Fig. \ref{FigInteraction} we have visualized the coherent and dissipative interactions of the ground and excited states present in our system. It is evident that the dynamics of the excited states, which are subject to decay, is governed by both, unitary and dissipative time evolution. For the excited levels we can combine the real detunings of the levels with imaginary terms, that correspond to broadening by decay, to yield complex energy terms. The resulting non-Hermitian time evolution of the excited states is expressed compactly in terms of a non-Hermitian Hamiltonian 
\begin{eqnarray}
\label{EqHNH}
\hat{H}_{\rm NH} \equiv \hat{H}_{\rm e} - \frac{i}{2} \sum_k \hat{L}_k^{\dagger} \hat{L}_k.
\end{eqnarray}
Also referred to as the no-jump Hamiltonian in the language of the quantum jump formalism \cite{Molmer}, $\hat{H}_{\rm NH}$ combines the Hamiltonian of the decaying excited subspace $\hat{H}_{\rm e}$ with the jump terms of the master equation (\ref{EqFullMaster}). For the system at hand we find
\begin{eqnarray}
\label{EqHNH2}
\hat{H}_{\rm NH}=&\tilde{\Delta}_{0}\left(\ket{T_0}\bra{T_0}+\ket{S_0}\bra{S_0}\right) + \tilde{\Delta}_1 \left(\ket{T_1}\bra{T_1} + \ket{S_1}\bra{S_1} \right) + \nonumber \\ &+ \tilde{\delta}_0 \ket{00} \ket{1}\bra{1} \bra{00} + \tilde{\delta}_1 \ket{T} \ket{1}\bra{1} \bra{T} + \nonumber \\ &+ \tilde{\delta}_2 \ket{11} \ket{1}\bra{1} \bra{11} + \tilde{\delta}_{1} \ket{S} \ket{1}\bra{1} \bra{S} + \nonumber \\ &+ g \left(\ket{T_0}\bra{1}\bra{T} + \ket{S_0}\bra{1}\bra{S} + \ket{T_1}\bra{1}\bra{11} + H. c.\right).
\end{eqnarray}
Here, we have defined complex `energies' $\tilde{\Delta}_n \equiv \Delta_n-\frac{i\gamma}{2}$, $\tilde{\delta}_n \equiv \delta_n-\frac{i\kappa}{2}$, $\Delta_n \equiv \Delta+m \cdot \beta$, and $\delta_{n} \equiv \delta+m \cdot \beta$, with $m$ being the number of atoms in state $\ket{1}$ ($\Delta_0=\Delta$, $\delta_0=\delta$). As will become clear further below, it is useful to set $\Delta_{-1} \equiv \Delta_1$.\\
In doing so, the detunings of the excited states are conveniently combined with their decay widths to complex detunings, where both their real and imaginary parts govern the strengths of the transitions involving the excited states.

\subsection*{Effective Hamiltonian and Lindblad operators}
As can be recognized from Fig. \ref{FigInteraction}, the coherent and dissipative couplings of the excited states can be concatenated to effective second-order processes between the ground states. An example for such an effective unitary process is given by the transition $\ket{00} \overset{\Omega}{\rightarrow} \ket{T_0} \overset{\Omega}{\rightarrow} \ket{00}$, facilitated by the coherent driving $\hat{V}$ of strength $\Omega$, resulting in an effective shift of ground level $\ket{00}$. In case the coherent de-excitation is replaced by a decay an effective dissipative process is formed. Here, $\ket{00} \overset{\Omega}{\rightarrow} \ket{T_0} \overset{\gamma}{\rightarrow} \ket{S}$ is an example for effective decay from state $\ket{00}$ to $\ket{S}$, involving spontaneous emission of an atomic excitation at a rate of $\gamma$. In this manner, all available combinations of weak optical excitation, non-Hermitian coupling between the excited states and either subsequent weak optical de-excitation or decay bundle together to effective second-order processes between the ground states. We assume the optical excitation $\hat{V}$ to be a perturbation of the non-Hermitian evolution given of the excited levels by $\hat{H}_{\rm NH}$. Consequently, their population will always be much lower than the population of the ground levels. On these grounds, we can perform adiabatic elimination of the excited levels and restrict the dynamics to the ground states.\\
In Ref. \cite{EO} we present an effective operator formalism based on second-order perturbation theory and adiabatic elimination to reduce the evolution of an open system to effective unitary and dissipative processes between its ground states. Applying this method simplifies the complexity of the Liouvillian dynamics of Eq. (\ref{EqFullMaster}) considerably, resulting in an effective master equation in Lindblad form
\begin{eqnarray}
\label{EqEffMaster}
\dot{\rho}=i\left[\rho,\hat{H}_{\rm eff}\right]+\sum_k \hat{L}^{k}_{\rm eff} \rho (\hat{L}^{k}_{\rm eff})^\dagger -\frac{1}{2}\left((\hat{L}^{k}_{\rm eff})^\dagger\hat{L}^{k}_{\rm eff}\rho+\rho(\hat{L}^{k}_{\rm eff})^\dagger\hat{L}^{k}_{\rm eff}\right)
\end{eqnarray}
represented by an effective Liouvillian $\mathcal{L}_{\rm eff}[\rho]=\dot{\rho}$. It contains an effective Hamiltonian $\hat{H}_{\rm eff}$ and effective Lindblad operators $\hat{L}_{\rm eff}^k$
\begin{eqnarray}
\label{EqEO}
\hat{H}_{\rm eff} &\equiv -\frac{1}{2} \hat{V}_{-}\left(\hat{H}_{\rm NH}^{-1}+(\hat{H}_{\rm NH}^{-1})^{\dagger}\right) \hat{V}_{+} + \hat{H}_{\rm g}\\
\label{EqEO2}
\hat{L}^{k}_{\rm eff} &\equiv \hat{L}_k \hat{H}_{\rm NH}^{-1} \hat{V}_{+}
\end{eqnarray}
incorporating the inverse of the non-Hermitian Hamiltonian $\hat{H}_{\rm NH}$ of Eq. (\ref{EqHNH}) and the unperturbed ground-state Hamiltonian $\hat{H}_{\rm g}$.\\
As expected, these effective processes consist of an initial weak optical excitation $\hat{V}_+$ and a final de-excitation $\hat{V}_-$ or decay $\hat{L}_k$ depending on their unitary or dissipative character. In-between, the inverse non-Hermitian Hamiltonian $\hat{H}_{\rm NH}^{-1}$ acts as a `propagator', representing the non-Hermitian evolution of the excited subspace. We find that its elements determine the strength of the effective process depending on which excited states take part in it. Its properties will be addressed in more detail in the following section.
\begin{figure*}[t]
\centering
\includegraphics[width=14cm]{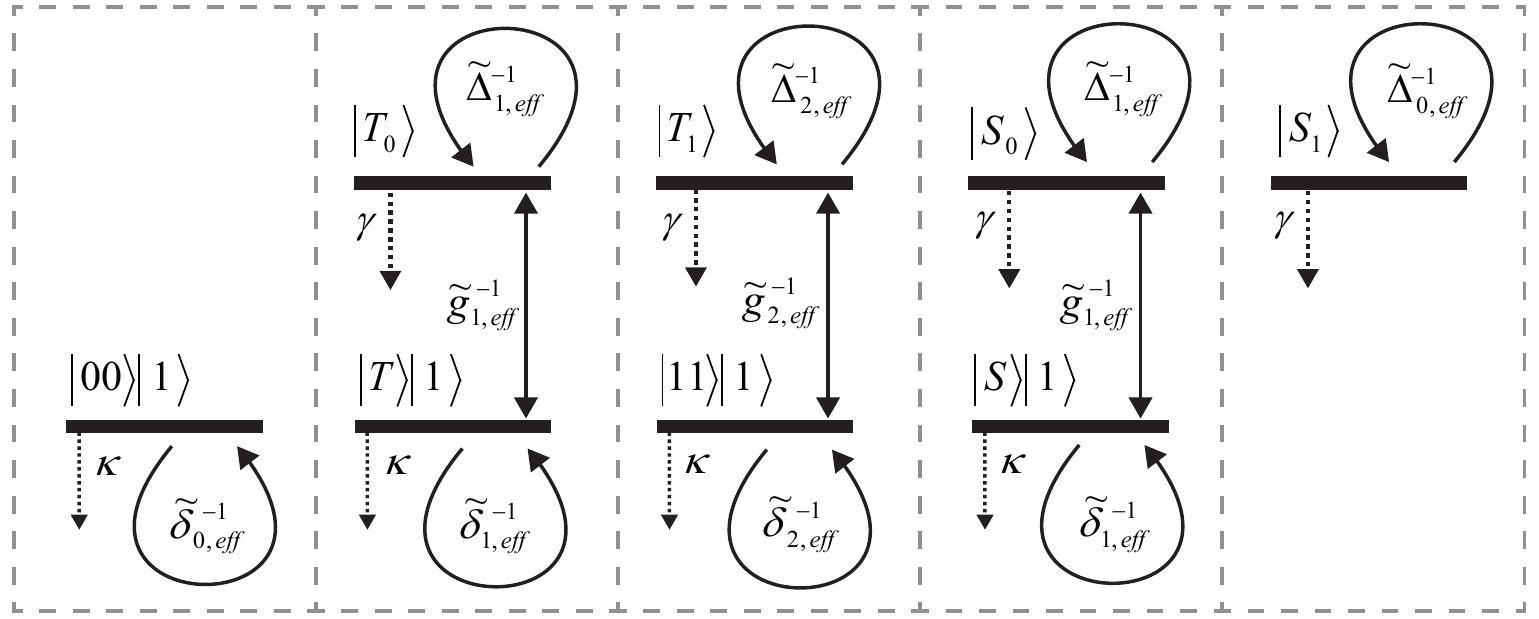}
\caption{Propagators in the excited-state subspace. The loop-like elements $\tilde{\Delta}_{n, \rm eff}^{-1}$, $\tilde{\delta}_{n, \rm eff}^{-1}$ and transition-like elements $\tilde{g}_{n, \rm eff}^{-1}$ contained in $\hat{H}_{\rm NH}^{-1}$ govern the non-Hermitian evolution of the excited states. Grouped according to the three interacting and two non-interacting excited subspaces these propagators determine the strength of effective processes involving the excited states, depending on the state reached by initial excitation and the one left by either coherent de-excitation or decay.}
\label{FigPropagators}
\end{figure*}

\subsection*{Effective propagators of the excited states}
In the excited-state basis defined earlier, $\hat{H}_{\rm NH}$ can be broken up into 5 block diagonal elements which evolve independently. $\hat{H}_{\rm NH}^{-1}$ is then also in block diagonal form, and can be written out explicitly as 
\begin{eqnarray}
\hat{H}_{\rm NH}^{-1} = \hat{H}_{{\rm NH},\ket{T_0}}^{-1} + \hat{H}_{{\rm NH},\ket{S_0}}^{-1} + \hat{H}_{{\rm NH},\ket{T_1}}^{-1} + \hat{H}_{{\rm NH},\ket{S_1}}^{-1} + \hat{H}_{{\rm NH},\ket{00}\ket{1}}^{-1}
\end{eqnarray}
with three blocks for the interacting excited subspaces
\begin{eqnarray}
\label{EqInteracting1}
\hat{H}_{{\rm NH},\ket{T_0}}^{-1} &= \tilde{\Delta}_{1, \rm eff}^{-1} \ket{T_0}\bra{T_0} + \tilde{\delta}_{1, \rm eff}^{-1} \ket{T}\ket{1}\bra{1}\bra{T} + \tilde{g}_{1, \rm eff}^{-1} \left(\ket{T}\ket{1}\bra{T_0} + H.c. \right) \\
\label{EqInteracting2}
\hat{H}_{{\rm NH},\ket{S_0}}^{-1} &= \tilde{\Delta}_{1, \rm eff}^{-1} \ket{S_0}\bra{S_0} + \tilde{\delta}_{1, \rm eff}^{-1} \ket{S}\ket{1}\bra{1}\bra{S} + \tilde{g}_{1, \rm eff}^{-1} \left(\ket{S}\ket{1}\bra{S_0} + H.c. \right) \\
\label{EqInteracting3}
\hat{H}_{{\rm NH},\ket{T_1}}^{-1} &= \tilde{\Delta}_{2, \rm eff}^{-1} \ket{T_1}\bra{T_1} + \tilde{\delta}_{2, \rm eff}^{-1} \ket{11}\ket{1}\bra{1}\bra{11} + \tilde{g}_{2, \rm eff}^{-1} \left(\ket{11}\ket{1}\bra{T_1} + H. c. \right) \nonumber\\
\end{eqnarray}
and two blocks for the non-interacting excited states
\begin{eqnarray}
\label{EqS1Subspace}
\hat{H}_{{\rm NH},\ket{S_1}}^{-1} &= \tilde{\Delta}_{0, \rm eff}^{-1} \ket{S_1}\bra{S_1}
\\
\hat{H}_{{\rm NH},\ket{00}\ket{1}}^{-1} &= \tilde{\delta}_{0, \rm eff}^{-1} \ket{00}\ket{1}\bra{1}\bra{00}.
\end{eqnarray}
In order to keep the notation compact we have defined here
\begin{eqnarray}
\label{EqDefR}
\tilde{\Delta}_{n, \rm eff} &\equiv 
\tilde{\Delta}_{n-1} - \frac{n g^2}{\tilde{\delta}_n}\\
\label{EqDefr}
\tilde{\delta}_{n, \rm eff} &\equiv 
\tilde{\delta}_n - \frac{n g^2}{\tilde{\Delta}_{n-1}}\\
\label{EqDefG}
\tilde{g}_{n, \rm eff} &\equiv 
\sqrt{n} g - \frac{\tilde{\delta}_n \cdot \tilde{\Delta}_{n-1}}{\sqrt{n} g},
\end{eqnarray}
The entries of $\hat{H}_{\rm NH}^{-1}$, shown in Fig. \ref{FigPropagators}, are generally complex and their magnitudes have dimension of inverse energy. They play the role of propagators or complex magnitudes in the effective ground state to ground state processes of Eqs. (\ref{EqEO}-\ref{EqEO2}). Each effective process that is formed from perturbative optical excitation $\hat{V}_+$ and subsequent de-excitation $\hat{V}_-$ or decay includes a propagator depending on which excited states are involved.\\
As can be seen in Fig. \ref{FigPropagators}, $\hat{H}_{\rm NH}^{-1}$ contains both loop-like and transition-like propagators, grouped according to the five separable subspaces. Their index $n$ reflects the coupling strength between the atomic and cavity excited state of the interacting subspace, the latter of which has $n$ atoms in state $\ket{1}$. The states $\ket{S_1}$ and $\ket{00}\ket{1}$ are dark-states of the atom-cavity interaction and are uncoupled ($n=0$).\\
By our definitions of Eqs. (\ref{EqDefR}-\ref{EqDefr}) we have associated the loop-like propagators $\tilde{\Delta}_{n, \rm eff}^{-1}$ and $\tilde{\delta}_{n, \rm eff}^{-1}$ with the complex detunings of the excited states such that for a vanishing coupling $g$ the shifts in $\tilde{\Delta}_{n, \rm eff}$ and $\tilde{\delta}_{n, \rm eff}$ would disappear, and we would find $\tilde{\Delta}_{n, \rm eff} = \tilde{\Delta}_{n-1}$ and $\tilde{\delta}_{n, \rm eff} = \tilde{\delta}_{n}$. Similarly, in case of negligible complex detunings $\tilde{\Delta}_{n-1}$ and $\tilde{\delta}_{n}$ the transition-like propagators $\tilde{g}_{n, \rm eff}^{-1}$ in Eqs. (\ref{EqInteracting1}-\ref{EqInteracting3}) would simplify to the inverse of a real coupling $\tilde{g}_{n, \rm eff} = \sqrt{n}g$, the well-known dressed state energy for $n$ atoms resonant with a cavity.\\
All propagators of $\hat{H}_{\rm NH}$ and Fig. \ref{FigPropagators} can be written in terms of a denominator $\tilde{D}_n \equiv n g^2 - \tilde{\delta}_{n} \cdot \tilde{\Delta}_{n-1}$ which equals the reduced determinant of the Hamiltonian of the according subspace and is highly dependent on the system parameters. As we will show, an appropriate choice of the parameters $\Delta$, $\delta$ and $\beta$ engineers certain propagators of the excited states, and hence, effective decay processes mediated by them, to be very strong, while others are effectively suppressed. Physically this can be understood as shifting one of the dressed states into resonance to enhance the coupling. In the following discussion we present various schemes that build upon this principle of engineered decay. Here, each of the schemes is denoted by the atomic excited state that mediates the engineered decay into the desired maximally entangled singlet state $\ket{S}$.\\
Applying the discussed methods we will be able to analytically derive the optimal parameters and benchmarks for each of the presented schemes, the most important of which are the steady-state fidelity with the desired entangled state and the convergence time, estimated by the inverse of the spectral gap (as explained later the convergence time can be estimated by the spectral gap which is the smallest non-zero eigenvalue of the Liouvillian). We back up all of our analytic results by numerical integration of the master equations (\ref{EqFullMaster}) and (\ref{EqEffMaster}).

\section[A scheme for engineered spontaneous emission]{A scheme for engineered spontaneous emission mediated by a dark state}
\label{SectionS1}
In the following we present a scheme for the preparation of an entangled steady state by an engineered spontaneous emission process mediated by the dark state of the atom-cavity interaction, $\ket{S_1}$. Of the considered schemes it exhibits the lowest error in the preparation of the entangled state.\\
As for all schemes discussed in this article, we begin by outlining the physical mechanisms that underlie the dissipative state preparation, and proceed with a discussion of the effective operators. Depending on the driving regime we analytically derive the benchmarks for each scheme, in particular steady-state fidelity and speed of convergence, from the effective dynamics.\\
For the scheme at hand, we choose to engineer the effective decay by spontaneous emission into the maximally entangled singlet state to be as strong as possible. To this end we set both the optical driving and the cavity transition to be resonant ($\Delta=0$, $\delta=-\beta$) and the microwave or Raman transition to be slightly detuned ($\beta \neq 0$). Furthermore, we set $\phi=\pi$ so that the optical driving crosses the singlet/triplet subspace, i. e. $\ket{T} \overset{\Omega}{\leftrightarrow} \ket{S_1}$ and $\ket{S} \overset{\Omega}{\leftrightarrow} \ket{T_1}$. 
\begin{figure*}[ht]
\centering
\includegraphics[width=11cm]{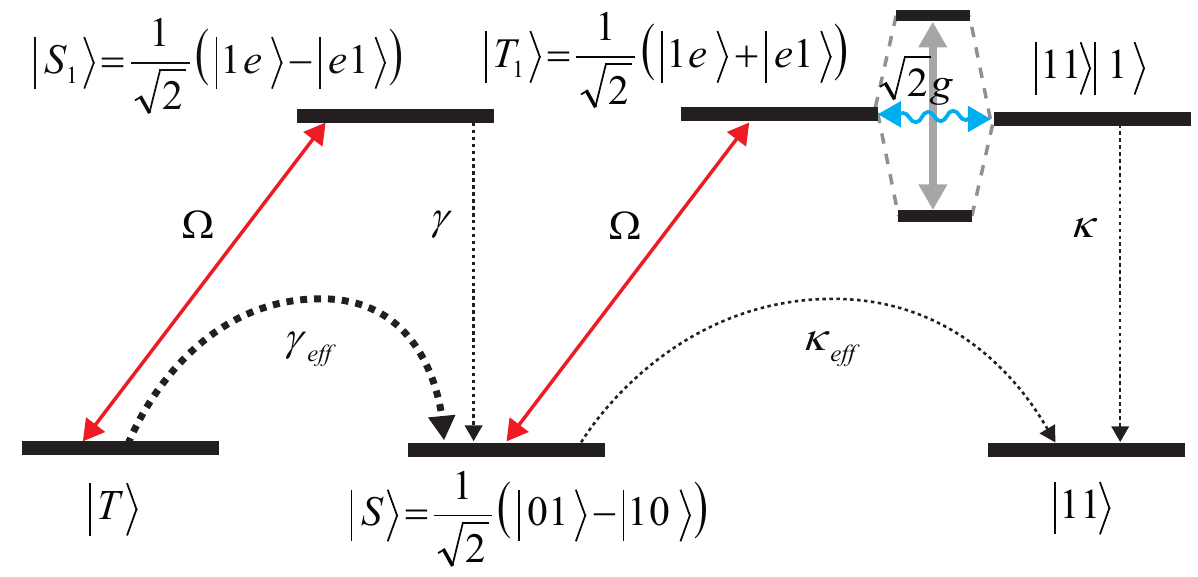}
\caption{Effective decay processes of the $\ket{S_1}$ scheme. The optical excitation $\Omega$ drives population from $\ket{T}$ to $\ket{S_1}$. From there it decays spontaneously into the desired steady state $\ket{S}$ with a certain probability. As $\ket{S_1}$ is the dark state of the atom-cavity interaction $\hat{H}_{\rm ac}$ the initial excitation is not shifted and is close to resonance so that the effective decay $\gamma_{\rm eff}$ from $\ket{T}$ prepares $\ket{S}$ very rapidly. As $\hat{H}_{\rm ac}$ strongly couples $\ket{T_1}$ and $\ket{11}\ket{1}$ with a strength $\sqrt{2} g$ these states form dressed states that are shifted out of resonance. Effective spontaneous emission $\gamma_{\rm eff}$ and cavity loss $\kappa_{\rm eff}$ from $\ket{S}$ into $\ket{11}$, mediated by $\ket{T_1}$, is hence effectively suppressed. A microwave/Raman transition (not shown) couples $\ket{00}$ and $\ket{11}$ to $\ket{T}$.}
\label{FigS1mechanism}
\end{figure*}

\subsection*{Mechanism of the state preparation}
\begin{figure*}[ht]
\centering
\includegraphics[width=14cm]{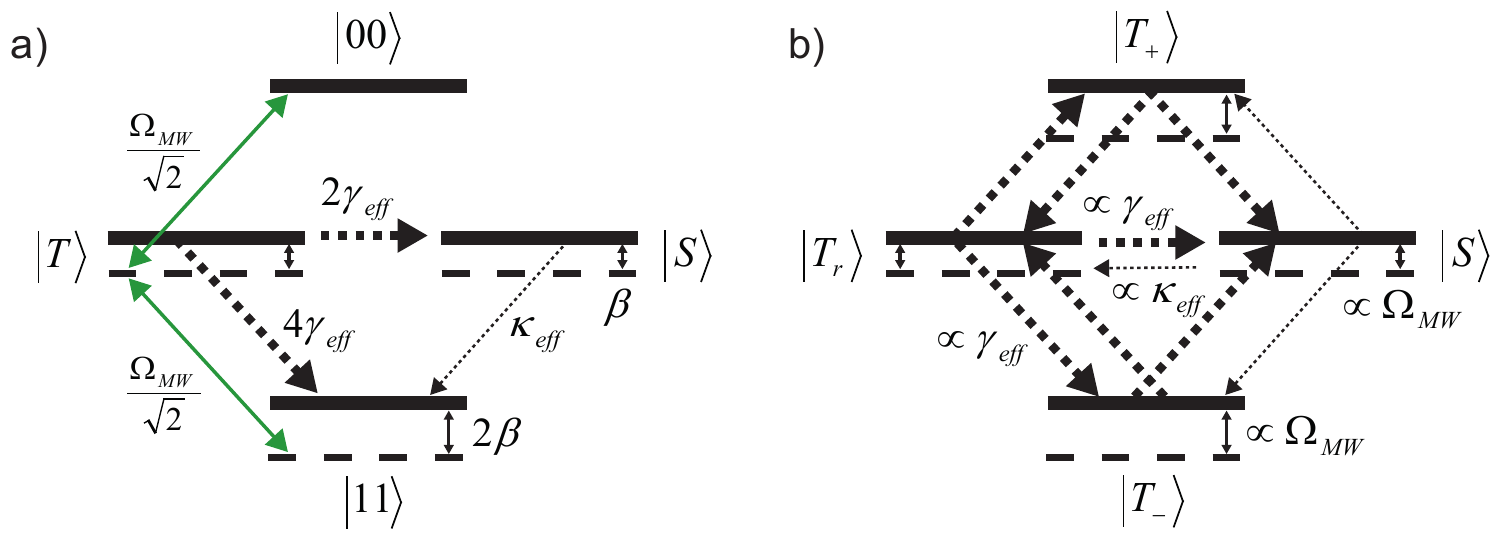}
\caption{Effective ground state processes of the $\ket{S_1}$ scheme. (a) Detuning and interactions in the shuffling picture, where $\ket{00}$, $\ket{T}$ and $\ket{11}$ are coupled by $\Omega_{\rm MW}$ to avoid population in $\ket{00}$ and $\ket{11}$. Engineered spontaneous emission prepares the maximally entangled singlet state $\ket{S}$ at a rate of $2\gamma_{\rm eff}$. Effective cavity decay out of $\ket{S}$ happens at a rate of $\kappa_{\rm eff}$. (b) Dressed state picture. Strong spontaneous emission $\propto \gamma_{\rm eff}$ reshuffles the dressed triplet states (not shown between $\ket{T_\pm}$). Population in $\ket{S}$ is gained from ($\propto \gamma_{\rm eff}$) and lost to ($\propto \kappa_{\rm eff}$) each of the dressed triplet states.}
\label{FigS1effective}
\end{figure*}

The mechanism is illustrated in Fig. \ref{FigS1mechanism}. Population from $\ket{T}$ is excited up to $\ket{S_1}=\frac{1}{\sqrt{2}}\left(\ket{1e}-\ket{e1}\right)$. The atomic excited state $\ket{S_1}$ is the dark state of the atom-cavity interaction, $\hat{H}_{\rm ac}$, and is therefore resonant with the optical driving ($\Delta=0$). Consequently, $\ket{T}$ decays very rapidly over $\ket{S_1}$ into $\ket{S}$. On the other hand, population from $\ket{S}$ is excited to $\ket{T_1}$ which is coupled to $\ket{T}\ket{1}$ with a strength of $\sqrt{2}g$. This strong coupling shifts their dressed states out of resonance by $\pm \sqrt{2}g$ which is much more than the natural linewidth. Decay out of $\ket{S}$ is thus strongly suppressed, while $\ket{T}$ is rapidly pumped into $\ket{S}$. Accumulation of population in $\ket{00}$ or $\ket{11}$ is prevented by the microwave/Raman field that couples the three triplet states $\ket{00}$, $\ket{T}$ and $\ket{11}$ and reshuffles population to $\ket{T}$ from which engineered decay prepares $\ket{S}$ again. The detuning $\beta$ of the microwave is needed to prevent $\frac{1}{\sqrt{2}}(\ket{00}-\ket{11})$ from being a dark state of the microwave which would not be reshuffled to $\ket{T}$. The effective processes resulting from the coupling are illustrated in Fig. \ref{FigS1effective} a). We note that even though the state $\ket{T_1}$ is far out of resonance the desired steady state $\ket{S}$ is still weakly coupled to $\ket{T_1}$ by the laser driving; $\ket{S}$ is hence not an ideal dark state. The fidelity of the steady state with $\ket{S}$ and the error rate of the protocol depend on the ratio of the rate of the dissipative preparation of $\ket{S}$ and the rate of decay from $\ket{S}$. In the following sections we will model these processes quantitatively by considering the effective operators to derive the optimal parameters and the error of the protocol analytically.

\subsection*{Effective processes}
We begin our discussion of the effective processes, shown in Fig. \ref{FigS1effective} a), by deriving the general effective operators for optical driving with $\phi=\pi$. Given $\hat{V}$ and $\hat{L}_k$ the terms for the effective processes can be read off directly from the map of propagators in Fig. \ref{FigPropagators}. These operators are equally valid for the $\ket{S_0}$ scheme in Sec. \ref{AppS0} that also uses $\phi=\pi$. For the effective decay of an atomic excitation $\ket{e}$ into ground state $\ket{0}$ by spontaneous emission we obtain the effective Lindblad operators
\begin{eqnarray}
\hat{L}_{\rm eff}^{\gamma,0,\{1,2\}} = &\pm \frac{\sqrt{\gamma/2} \Omega}{2 \tilde{\Delta}_{1, \rm eff}} \ket{00}\bra{00} + \frac{\sqrt{\gamma/2} \Omega}{4 \tilde{\Delta}_{0, \rm eff}} \left(\pm \ket{T}\bra{T} + \ket{S}\bra{T}\right) + \nonumber \\ & + \frac{\sqrt{\gamma/2} \Omega}{4 \tilde{\Delta}_{2, \rm eff}} \left(\ket{T}\bra{S} \pm \ket{S}\bra{S} \right).
\end{eqnarray}
The superscript $\gamma$ on the Lindblad operators stands for spontaneous emission, $0$ for the decay into ground state $\ket{0}$, and the index $\{1,2\}$ for the atom at which the decay occurs refers to the upper (lower) set of signs of the terms. Similarly, the effective decay by spontaneous emission into ground state $\ket{1}$ is given by
\begin{eqnarray}
\hat{L}_{\rm eff}^{\gamma,1,\{1,2\}} = &+ \frac{\sqrt{\gamma} \Omega}{4 \tilde{\Delta}_{1, \rm eff}} \left(\pm\ket{T}\bra{00}-\ket{S}\bra{00}\right) \pm \frac{\sqrt{\gamma} \Omega}{4 \tilde{\Delta}_{0, \rm eff}} \ket{11}\bra{T} + \nonumber \\
& + \frac{\sqrt{\gamma} \Omega}{4 \tilde{\Delta}_{2, \rm eff}} \ket{11}\bra{S}.
\end{eqnarray}
The effective decay of a cavity excitation is found to give
\begin{eqnarray}
\hat{L}_{\rm eff}^\kappa = \frac{\sqrt{\kappa} \Omega}{2 \tilde{g}_{2, \rm eff}} \ket{11}\bra{S} - \frac{\sqrt{\kappa/2} \Omega}{\tilde{g}_{1, \rm eff}} \ket{S}\bra{00}.
\end{eqnarray}
Finally, the effective unitary processes are given by
\begin{eqnarray}
\label{EqS1Heff}
\hat{H}_{\rm eff} = &-{\rm Re} \left[\frac{\Omega^2}{2 \tilde{\Delta}_{1, \rm eff}}\right] \ket{00}\bra{00} 
- {\rm Re} \left[\frac{\Omega^2}{4 \tilde{\Delta}_{0, \rm eff}}\right] \ket{T}\bra{T} -  \nonumber \\ &- {\rm Re} \left[\frac{\Omega^2}{4 \tilde{\Delta}_{2, \rm eff}}\right] \ket{S}\bra{S} + \hat{H}_{\rm g}.
\end{eqnarray}
where ${\rm Re}[ \ ]$ denotes the real part of the argument.
While the above effective operators hold whenever $\phi=\pi$, we can simplify them for the particular scheme at hand by discussing the propagators for the parameter choices made in the previous section. In the absence of an atomic detuning, $\Delta=0$, the complex energy of $\ket{S_1}$, as the dark-state of the cavity interaction, is given by $\tilde{\Delta}_1 = \beta-\frac{i\gamma}{2}$. As we will discuss below it is desirable to have $\beta \propto \Omega$ so that for the assumption of weak driving ($\Omega \ll \gamma$) we can write $\tilde{\Delta}_1 \approx -\frac{i\gamma}{2}$. The propagator of the effective $\ket{S_1}$-mediated decay processes then simplifies to
\begin{eqnarray}
\bra{S_1}\hat{H}_{\rm NH}^{-1}\ket{S_1}=\tilde{\Delta}_{0, \rm eff}^{-1} \approx - \frac{2}{i\gamma}.
\end{eqnarray}
Hence, the effective decay processes mediated by $\ket{S_1}$ that incorporate this propagator are tailored to be very strong compared to the decay out of the singlet state which involves the subspace consisting of the states $\ket{T_1}$ and $\ket{11}\ket{1}$, and the transition-like propagator
\begin{eqnarray}
\bra{1}\bra{T}\hat{H}_{\rm NH}^{-1}\ket{T_1} = \tilde{g}_{2, \rm eff}^{-1} \approx \frac{1}{\sqrt{2} g}.
\end{eqnarray}
The last denominator reflects the strong shift of the dressed states of $\ket{T_1}$ and $\ket{11}\ket{1}$ out of resonance, slowing down the effective decay out of $\ket{S}$. Consequently, we have reached $|\tilde{\Delta}_{0, \rm eff}^{-1}| \gg |\tilde{g}_{n, \rm eff}^{-1}|$ so that effective processes mediated by the dark state $\ket{S_1}$ are engineered to be much stronger than those involving other states, in particular $\ket{T_1}$. We have thus found that the triplet ground state $\ket{T}$ undergoes rapid effective spontaneous emission at a rate $\propto 1/\gamma$, while cavity decay from $\ket{S}$ $\propto \kappa/g^2$ is suppressed in the strong coupling regime where $C \gg 1$. For finding the steady state we can drop the suppressed terms unless they affect the singlet state. In addition, the spontaneous emission from $\ket{S}$ $\propto \gamma\kappa^2/g^4$ is negligible and will be ignored. The effective decay processes then simplify to
\begin{eqnarray}
\label{EqS1EO}
\hat{L}_{\rm eff}^{\gamma,0,\{1,2\}} &= \pm i \sqrt{\gamma_{\rm eff}} \ket{T}\bra{T} + i \sqrt{\gamma_{\rm eff}} \ket{S}\bra{T}\\
\hat{L}_{\rm eff}^{\gamma,1,\{1,2\}} &= \pm i \sqrt{2 \gamma_{\rm eff}} \ket{11}\bra{T}\\
\label{EqS1EO3}
\hat{L}_{\rm eff}^\kappa &= \sqrt{\kappa_{\rm eff}} \ket{11}\bra{S}.
\end{eqnarray}
Here, we have set $\kappa_{\rm eff} = |\bra{11}\hat{L}_{\rm eff}^{\kappa}\ket{S}|^2 = \frac{\kappa \Omega^2}{8 g^2}$ and $\gamma_{\rm eff} =|\bra{S}\hat{L}_{\rm eff}^{\gamma, 0,\{1,2\}}\ket{T}|^2 = \frac{\Omega^2}{8 \gamma}$.
Furthermore, for the scheme at hand the effective Hamiltonian $\hat{H}_{\rm eff}$ of Eq. (\ref{EqS1Heff}) is well approximated by the unperturbed ground-state Hamiltonian $\hat{H}_{\rm g}$
\begin{eqnarray}
\hat{H}_{g} = &\frac{\Omega_{\rm MW}}{2} \left(\ket{00}\bra{T} + \ket{T}\bra{11} + H. c.\right) + \beta \left(2\ket{11}\bra{11} + \ket{T}\bra{T} + \ket{S}\bra{S} \right), \nonumber
\end{eqnarray}
where we have neglected the minor effective shifts $\mathcal{O}(\Omega^2)$.
The resulting effective decay processes of this scheme are illustrated in Fig. \ref{FigS1effective} a) together with the microwave/Raman reshuffling. The singlet state $\ket{S}$ is efficiently prepared from $\ket{T}$ by spontaneous emission at a rate of $2\gamma_{\rm eff}$. The singlet $\ket{S}$ decays by effective cavity loss $\kappa_{\rm eff}$ into $\ket{11}$. The mechanism that allows us to engineer a strong effective spontaneous emission from $\ket{T}$ into $\ket{S}$ at the same time causes strong decay at a rate of $4 \gamma_{\rm eff}$ from $\ket{T}$ into $\ket{11}$. Hence, accumulation in $\ket{11}$ needs to be inhibited by the microwave/Raman shuffling $\Omega_{\rm MW}$. The fidelity of the steady state with the desired entangled state will be be derived analytically in the following sections after changing into a dressed state picture with respect to $\Omega_{\rm MW}$.

\subsection*{Parameter analysis at weak driving}
We first analyze the dynamics of this $\ket{S_1}$ scheme for weak optical driving $\Omega$. After a basis transform into a dressed ground state picture, this assumption allows us to reduce the dynamics to rate equations for the ground state populations. From these, we derive the important benchmarks for a comparison of the presented schemes; the steady-state fidelity with the desired entangled state, and the spectral gap as a measure for the rate of convergence.\\
\\
The basis used so far, involving the triplet states $\ket{00}$, $\ket{T}$ and $\ket{11}$ coupled by $\Omega_{\rm MW}$, will be referred to as `shuffling picture' in the remainder of the paper. We now simplify the analysis by identifying a basis in which non-diagonal elements of the density matrix of the reduced system are suppressed, as a consequence of the weak driving. It is then possible to express the dynamics as a set of linear rate equations. The basis of the new `dressed state picture' contains the original singlet state $\ket{S}$, and the three dressed triplet states
\begin{eqnarray}
\ket{T_\pm}=-1/2 (B \mp 1) \ket{00} + 1/2 (B \pm 1) \ket{11}+ A/\sqrt{2} \ket{T} \\
\ket{T_r}= A/\sqrt{2} \ket{00} - A/\sqrt{2} \ket{11}+ B \ket{T},
\end{eqnarray}
where we have defined $A=\Omega_{\rm MW}/\sqrt{\Omega_{\rm MW}^2+\beta^2}$ and $B=\beta/\sqrt{\Omega_{\rm MW}^2+\beta^2}$. In this basis the effective Hamiltonian is diagonal
\begin{eqnarray}
\hat{H}_{\rm eff} = &\sum_{+/-} \left(\beta \pm \left(B \beta + A \Omega_{\rm MW}\right)\right) \ket{T_\pm}\bra{T_\pm} + \beta \left(\ket{T_r}\bra{T_r} + \ket{S}\bra{S}\right).
\end{eqnarray}
The parameters $A$ and $B$, and hence the ratio of $\beta$ to $\Omega_{\rm MW}$, determine the contribution of $\ket{T}$ to each of the dressed states. We find that the optimal fidelity is obtained at $A=\sqrt{\frac{2}{3}}$, $B=\sqrt{\frac{1}{3}}$, and $\beta=\Omega_{\rm MW}/\sqrt{2}$. Here, each of the dressed states contains an equal share of the triplet state $\ket{T}$ from which $\ket{S}$ is prepared.\\
In the weak driving regime ($\Omega \ll \gamma$), the rephasing of the dressed states is much faster than the effective decay $\gamma_{\rm eff} \propto \frac{\Omega^2}{\gamma}$. Consequently, in the new basis the evolution of the coherences can be dropped from the master equation. The dissipative time evolution is then well approximated by a set of coupled linear differential rate equations. The rate equation for the population of the singlet state $P_S$ is 
\begin{eqnarray}
\label{EqRateS1}
\dot{P}_{S} &= \frac{\Omega^2}{12 \gamma} \left(P_{T_+} + P_{T_-} + P_{T_r}\right) - \frac{\kappa \Omega^2}{8 g^2} P_{S}.
\end{eqnarray}
Here, we have used that for weak driving the decay from the three dressed triplet states into the singlet are of the same strength due to the equal weight of $\ket{T}$ in the dressed states and equals one third of the total rate $\Omega/4\gamma$.

\subsection*{Derivation of the static error}
\label{Weak}
From the rate equation (\ref{EqRateS1}) we can derive the fidelity of the steady state with respect to the maximally entangled singlet state as $F_S=\underset{t \rightarrow \infty}{\lim} \ P_S$. Equivalently, the error of the protocol is found as the stationary population of the undesired triplet states $\left(1-F_S\right)$. Inserting $1-P_S=P_{T_+}+P_{T_-}+P_{T_r}$ into Eq. (\ref{EqRateS1}), we use $\dot{P}_S=0$ and obtain for the static error of the protocol
\begin{eqnarray}
\label{EqStaticError}
\left(1-F_S\right)_{\rm stat} = \frac{3 \gamma \kappa}{2 g^2} \equiv \frac{3}{2C},
\end{eqnarray}
with the cooperativity $C$ as defined in Sec. \ref{SectionSetup}. Eq. (\ref{EqStaticError}) indicates that in the strong coupling regime ($C \gg 1$) the only non-negligible error term scales linear in $C^{-1}$. This linear scaling of the error in the cooperativity is similar to Ref. \cite{KRS}, but the constant pre-factor is improved from $\frac{7}{2}$ to $\frac{3}{2}$, which could be important for experimental realizations, as discussed in the comparison section \ref{SectionComparison}.  
\begin{figure*}[t]
\centering
\includegraphics[width=15cm]{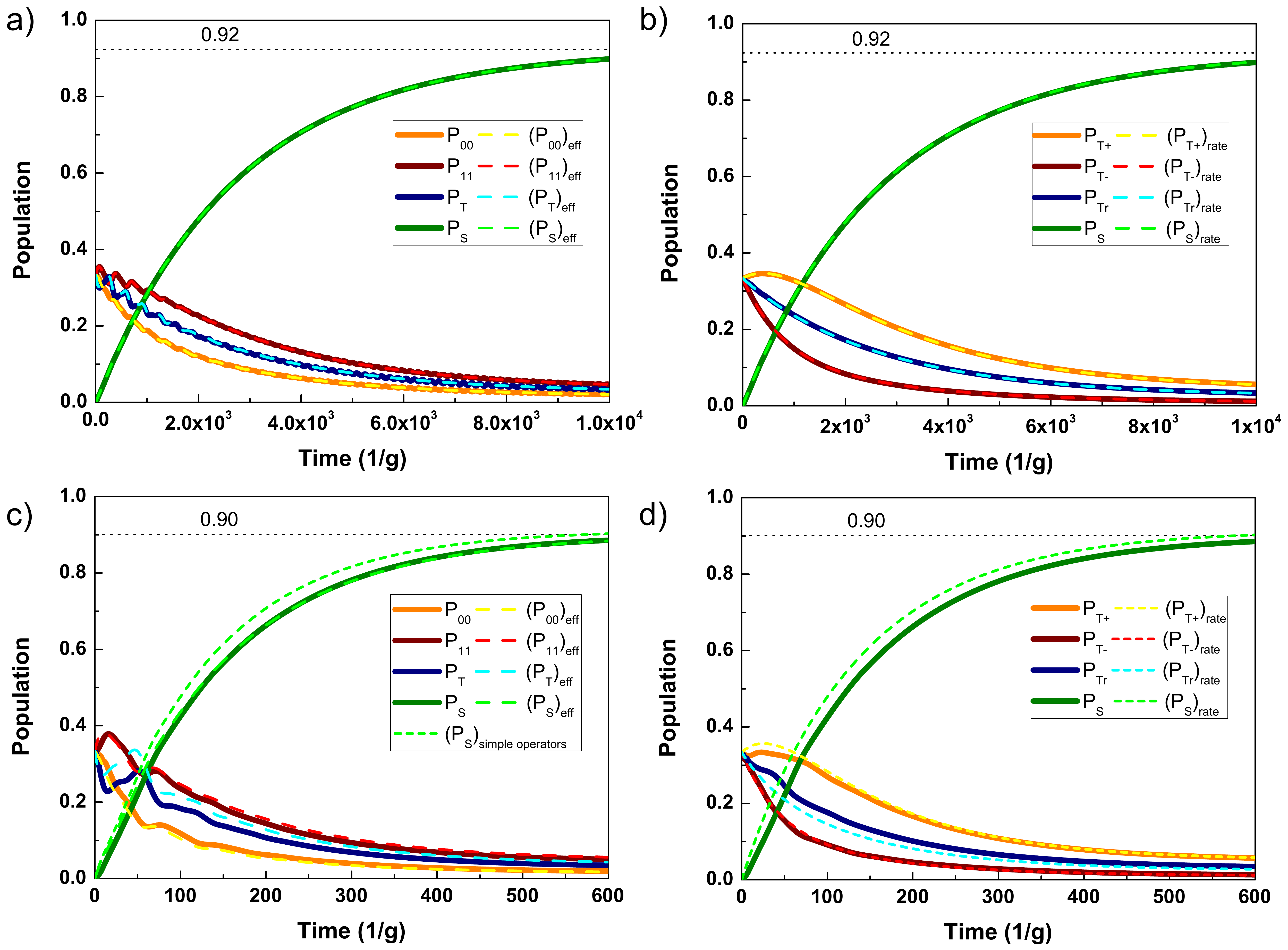}
\caption{Evolution of the system towards the entangled steady state for $\left(\gamma,\kappa\right)=\left(\frac{3g}{8},\frac{5g}{32}\right)$ similar to Ref. \cite{Kubanek1}, corresponding to $C \approx 17$. Dynamics of the full master equation (solid lines) are compared with effective dynamics in the shuffling picture (a, c) and rate equations in the dressed state picture (b, d) (dashed) for weak-driving (a-b) and increased driving (c-d). Starting from a completely mixed triplet state (see legend for details) the system evolves towards the maximally entangled singlet state ($\ket{S}$ -- green) approaching the steady-state fidelity (dotted line, indicated). For a weak driving of $\Omega=\frac{\gamma}{10}$ (a-b) the dynamics are completely described by rate equations of the populations, shown in (b). At increased driving $\Omega=\frac{\gamma}{2}$ (c-d) the dressed effective operators (long dash in c) are almost indistinguishable from the full dynamics, while simple effective operators and rate equations (short green dash in c), d) resp.) exhibit increasing inaccuracies. For all curves the optimized parameters $\Omega=2^{5/4}\Omega_{\rm MW}=2^{7/4}\beta$ (see also Sec. \ref{SectionSpeed}) were used.}
\label{FigS1Evolution}
\end{figure*}

\subsection*{Derivation of the spectral gap for weak-driving}
The quality of a dissipative state preparation protocol is determined by two main benchmarks: (i) fidelity of the stationary state, and (ii) speed of convergence of the protocol. We now consider the latter. Estimating the speed of convergence is in general a difficult task, but, for small systems, the spectral gap of the Liouvillian\footnote{The spectral gap of a Liouvillian $\mathcal{L}$ is defined as the magnitude of the smallest (in absolute value) non-zero real part of the eigenvalues of $\mathcal{L}$, where the Liouvillian is written as a linear operator in the matrix units basis, see Ref. \cite{WolfPerezGarcia} for further details.} is a very good estimate of the rate of convergence. The spectral gap can be seen as the decay rate of the slowest-decaying quasi-stationary eigenstate. If the gap is small, then the quasi-stationary eigenstate remains populated for a long period of time, whereas if the gap is large, then all eigenstates except the stationary one get depopulated rapidly.\\
In the setting at hand, the gap can in fact be read off from the expressions for the effective decay process $\hat{L}^{\gamma,0,\{1,2\}}_{\rm eff}$, which have a rate $\gamma_{\rm eff}=\frac{\Omega^2}{8\gamma}$. As stated above, the dressed states each contain an equal share of $\frac{1}{\sqrt{3}} \ket{T}$. Hence, the singlet is prepared equally fast by decay of the three dressed states at an individual rate of $2 \cdot \frac{1}{3} \cdot \gamma_{\rm eff}$ which results in the spectral gap 
\begin{eqnarray}
\lambda = \frac{\Omega^2}{12 \gamma}.
\end{eqnarray}
Furthermore, $\lambda$ is recognized as the eigenvalue for the according lowest-lying eigenvector of the Liouvillian $P_\mathcal{T}\equiv\frac{1}{3}\left(P_{T_+}+P_{T_-}+P_{T_r}\right)$, as can be seen from Eq. (\ref{EqRateS1}) above.\\
\\
To confirm these predictions we have performed numerical simulation of the dynamics of the system. Fig. \ref{FigS1Evolution} (a-b) summarizes the results obtained by numerical integration of the master equation of the full system in Eq. (\ref{EqFullMaster}), consisting of ground states and singly excited states. These curves are plotted together either with those from the effective master equation in Eq. (\ref{EqEffMaster}) in (a), or from the rate equations (such as Eq. (\ref{EqRateS1}) for the singlet) in (b): For a weak driving ($\Omega=\frac{\gamma}{10}$) we compare the numerically obtained curves of the population dynamics of the full master equation (solid lines) with (a) the effective master equation in the shuffling picture and (b) the rate equations in the dressed state picture (both dashed). We see that in this regime the full and the effective dynamics of the system are in excellent agreement. In addition, the analytical quantities derived from the rate equations in this Section are found to describe the fidelity of the steady state with the maximally entangled singlet state and the convergence time very accurately.

\section{How fast can two atoms be entangled by dissipation?}
\label{SectionSpeed}
The fidelity of the prepared state with respect to the desired state gives us a measure of the quality of our scheme once the system has reached equilibrium. The second figure of merit of a dissipative state preparation protocol is the time required to reach convergence. In this section, we analyze how fast the scheme presented above can be performed. We emphasize in particular the trade-off between the speed of the protocol and its fidelity.\\
Speeding up the state preparation can be done by increasing the optical driving. Indeed, as can be seen from Fig. \ref{FigS1Evolution} (c, d), using an increased optical driving $\Omega$ improves the convergence time by several orders of magnitude at the expense of only a few per cent additional error. The main reason for the decrease in fidelity is that the strong driving requires a strong microwave shuffling of the population of the triplet ground states. The microwave field, in turn, shifts the ground states out of resonance. This results in a decrease of the fidelity at increased optical driving which we will investigate in detail below.\\
In order to model the effective dynamics of our scheme accurately even for increased optical driving, we begin this section by introducing an extended effective operator formalism to account for the coherent coupling of the ground states which has so far been ignored when adiabatically eliminating the excited states. We then proceed to analytically derive the scaling of the two main performance measures, error and convergence time (spectral gap), as a function of the strength of the coherent driving, and perform a study of the optimal preparation time of an entangled state of a given fidelity.

\subsection*{Effective processes in the presence of ground state dressing}
In the following section we will discuss an extension of the effective operator formalism presented in Section \ref{SectionMethods}. It allows us to include the effects of increased driving and dressing of the ground states and to derive the dynamic benchmarks of the scheme at hand.\\
So far, we have worked within the weak driving limit ($\Omega \ll \gamma, \kappa$), where simple perturbation theory holds very reliably. We now want to consider how our scheme behaves as we approach the increased driving regime. From the curves of Fig. \ref{FigS1Evolution} a) and b) we notice an excellent agreement between the dynamics simulated with the full and effective master equation, and the rate equations for a weak optical driving as low as $\Omega = \frac{\gamma}{10}$. On the other hand, for Fig. \ref{FigS1Evolution} c) and d), we have used $\Omega=\frac{\gamma}{2}$ which is clearly beyond the weak driving limit. Here, the previously employed simple effective operators and rate equations become increasingly inaccurate. This is due to the fact that in our derivation of Eqs. (\ref{EqEO}-\ref{EqEO2}) we have neglected the influence of the ground-state Hamiltonian $\hat{H}_{\rm g}$ on the effective processes. As we derive below, $\Omega_{\rm MW}$ has to be proportional to $\Omega$ so that an increased $\Omega$ also leads to a higher $\Omega_{\rm MW}$. For the case at hand, this influence induces a shift of the ground states by the microwave driving (`dressing'). In a frame where $\hat{H}_{\rm g}$ is diagonal we can include these effects by applying more general effective operators \cite{EO} 
\begin{eqnarray}
\hat{H}_{\rm eff}=-\frac{1}{2}\left[\hat{V}_{-} \sum_l \left(\hat{H}_{\rm NH}-E_l\right)^{-1} \hat{V}_{+}^l + H. c. \right]+ \hat{H}_{\rm g}\\
\hat{L}^{k}_{\rm eff}= \hat{L}_k \sum_l \left(\hat{H}_{\rm NH}-E_l\right)^{-1} \hat{V}_{+}^{l},
\end{eqnarray}
where $E_l$ is the energy of the initial ground state $l$ and $\hat{V}_{+}^{l}$ the excitation from it. For non-negligible ground state energies $E_l$, the elements of $\hat{H}_{\rm NH}^{-1}$ are generally replaced by $\hat{H}_{\rm NH}^{-1} \rightarrow (\hat{H}_{\rm NH}-E_l)^{-1}$. Yet, we note that in order to capture the effects of the dressed ground states it will not be necessary to keep the dressed state energies $E_l$ in all propagators, but only those engineered to be strong. In fact, numerical curves obtained from these extended operators match the evolution of the full master equation very accurately, as can be seen from Fig. \ref{FigS1Evolution} c).\\
\\
As soon as ground state dressing is taken into account, the decay rates from the dressed triplet states $\ket{T_\pm}$ and $\ket{T_r}$ into $\ket{S}$ are no longer identical. Effective spontaneous emission from the dressed triplet states into the singlet mediated by $\ket{S_1}$ exhibits the non-degenerate propagators
\begin{eqnarray}
\bra{S_1}\left(\hat{H}_{\rm NH}-E_{T_\pm}\right)^{-1}\ket{S_1} = \left(\tilde{\Delta}_{0, \rm eff}-E_{T_\pm}\right)^{-1} = \left(-\frac{i \gamma}{2}\mp\sqrt{\frac{3}{2}}\Omega_{\rm MW}\right)^{-1}\\
\bra{S_1}\left(\hat{H}_{\rm NH}-E_{T_r}\right)^{-1}\ket{S_1} = \left(\tilde{\Delta}_{0, \rm eff}-E_{T_r}\right)^{-1} = \left(-\frac{i \gamma}{2}\right)^{-1},
\end{eqnarray}
resulting in detuned decay rates $\gamma^{(T_\pm)}_{\rm eff}=\gamma_{\rm eff} \cdot \gamma^2/(\gamma^2+\Omega_{\rm MW}^2)$, while $\gamma^{(T_r)}_{\rm eff}=\gamma_{\rm eff}$ and $\kappa_{\rm eff}^{(S)} \approx \kappa_{\rm eff}$ are effectively unchanged. Taking these into account we can set up the rate equations in the same manner as in the weak driving case of Sec. \ref{Weak}.

\begin{figure*}[ht]
\centering
\includegraphics[width=15cm]{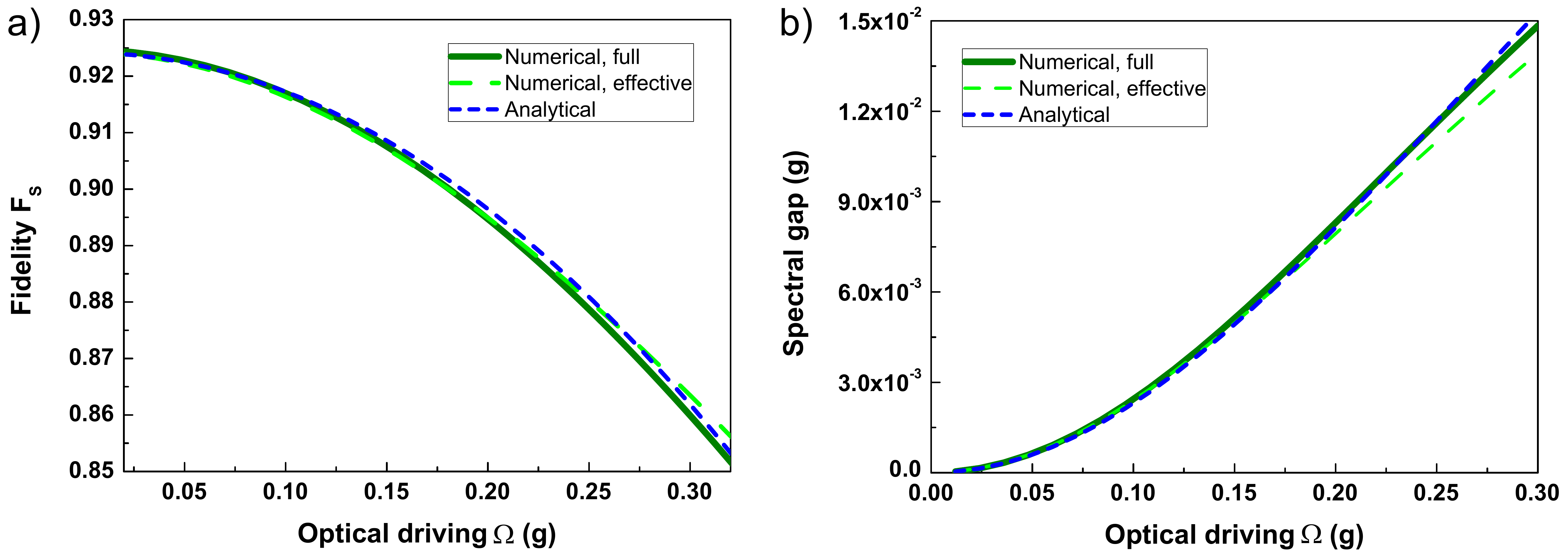}
\caption{Fidelity and spectral gap as a function of driving strength. (a) Fidelity of the steady state with the singlet state and (b) spectral gap as a measure of the speed of convergence with respect to the optical driving strength $\Omega$. Analytical results (blue, short dash) are in very good agreement with numerical curves obtained from the full (dark green) and effective (green dash) Liouvillian even at $\Omega \approx \gamma/2 \approx 0.2g$. For all curves the cavity parameters $(\gamma, \kappa)=(\frac{3g}{8},\frac{5g}{32})$ \cite{Kubanek1}, corresponding to $C \approx 17$, and the optimized driving parameters $\Omega=2^{5/4}\Omega_{\rm MW}=2^{7/4}\beta$ were used. Note that the analytical curve in a) contains terms that are not included in Eq. (\ref{EqDressingError2}) (see discussion of Eq. (\ref{EqError})).
}
\label{FigDriving}
\end{figure*}
\subsection*{Derivation of the error and of the spectral gap in the presence of ground state dressing}
\label{SectionDressingError}
Employing the state-dependent decay rates  $\gamma^{(T_l)}_{\rm eff}$ the additional error originating from the dressing of the triplet states is derived the following way:\\
Despite the different decay rates into the singlet, the population of the dressed triplet states is kept close to an equilibrium by strong dissipative shuffling $\propto 1/\gamma$ in-between them (see Fig. \ref{FigS1effective} b). Consequently, an equal mixture of the triplet states $P_{\mathcal{T}} \equiv \frac{1}{3} \left(P_{T_+}+P_{T_-}+P_{T_r}\right)$ is the slowest decaying eigenvector of the Liouvillian. Using this definition we set up the rate equation for the population of the singlet state
\begin{eqnarray}
\label{EqRateS1dressing}
\dot{P}_{S} &= \frac{\Omega^2}{12 \gamma} \frac{\gamma^2 + 2 \Omega_{\rm MW}^2}{\gamma^2 + 6 \Omega_{\rm MW}^2} P_{\mathcal{T}} - \frac{\kappa \Omega^2}{8 g^2} P_{S}.
\end{eqnarray}
While the loss of population from the singlet by cavity decay $\kappa_{\rm eff}$ is unaffected by the dressing, the decay rate of the triplet population through spontaneous emission has now become dependent on $\Omega_{\rm MW}$: Introducing $P_\mathcal{T}$ results in an effective decay rate of $\gamma^{(\mathcal{T})}_{\rm eff} \equiv \frac{3}{2} \cdot \eta \cdot \gamma_{\rm eff}$ with a factor $\eta \equiv \frac{\gamma^2 + 2 \Omega_{\rm MW}^2}{\gamma^2 + 6 \Omega_{\rm MW}^2}$ originating from averaging the decay rates $\gamma^{(T_l)}_{\rm eff}$. For the steady state ($\dot{P}_S=0$, $P_S \approx 1$) we derive the error
\begin{eqnarray}
\label{EqDressingError}
\left(1-F_S\right)&=\frac{3}{2 C} \cdot \frac{\gamma^2 + 6 \Omega_{\rm MW}^2}{\gamma^2 + 2 \Omega_{\rm MW}^2} \approx \frac{3}{2 C} \left(1+\frac{4 \Omega_{\rm MW}^2}{\gamma^2}\right) \\
&\equiv \left(1-F_S\right)_{\rm stat}+\left(1-F_S\right)_{\rm dres}.
\label{EqDressingError2}
\end{eqnarray}
As can be seen from the second step where we have expanded for $\Omega_{\rm MW} \ll \gamma$, the errors decouple into the static error, derived in Sec. \ref{Weak}, and another dynamic error $\left(1-F_S\right)_{\rm dres}$. The latter emerges from the dressing of the ground states by $\Omega_{\rm MW}$. Just as the static error, it decreases linearly with one over the cooperativity $C^{-1}$.\\
In the same manner, the spectral gap in the presence of ground state dressing is found from Eq. (\ref{EqRateS1dressing}),  determined by the decay rate of the lowest lying eigenvector $P_\mathcal{T}$,
\begin{eqnarray}
\label{EqDrivingGap4}
\lambda &= \frac{\Omega^2}{12 \gamma} \cdot \frac{\gamma^2 + 2 \Omega_{\rm MW}^2}{\gamma^2 + 6 \Omega_{\rm MW}^2}.
\end{eqnarray}
This result can also be derived more rigorously if we set up the full rate equations and extract the spectral gap as their smallest non-zero eigenvalue. We find
\begin{eqnarray}
\label{EqDrivingGap}
\lambda&=\frac{\Omega^2 \left(5 \gamma ^2+18 \Omega_{\rm MW}^2-\sqrt{9 \gamma ^4+84 \gamma ^2 \Omega_{\rm MW}^2+324 \Omega_{\rm MW}^4}\right)}{24 \gamma \left(\gamma^2+6 \Omega_{\rm MW}^2\right)} \\
&\approx
\frac{\Omega^2}{12 \gamma} \cdot \frac{\gamma^2 + 2 \Omega_{\rm MW}^2}{\gamma^2 + 6 \Omega_{\rm MW}^2}.
\label{EqDrivingGap2}
\end{eqnarray}
In the last line we have used $\gamma \gg \Omega_{\rm MW}$ and expanded up to second order in $\Omega_{\rm MW}$ which reproduces the result of Eq. (\ref{EqDrivingGap4}). For $\Omega_{\rm MW} \rightarrow 0$ the derived expressions reduce to the weak driving case as expected. In Fig. \ref{FigDriving} b) we have plotted the analytic result for the spectral gap (from Eq. (\ref{EqDrivingGap})) with respect to the optical driving $\Omega$, together with the numerically obtained spectral gap of the full and effective Liouvillians of Eqs. (\ref{EqFullMaster}) and (\ref{EqEffMaster}). We see that also for increased driving the curves are in good accordance.

\subsection*{Beyond rate equations}
So far, we have carried out our analytic study using rate equations formulated in a dressed state picture, where $\hat{H}_{\rm g}$ is diagonal. In the same picture, we have included the dressed ground state energies into the effective operators. However, in order to fully describe the system, in particular the effects at low microwave driving, we change back into the original `shuffling picture' with triplet states $\ket{00}$, $\ket{11}$ and $\ket{T}$ coupled by $\Omega_{\rm MW}/\sqrt{2}$. Introducing new decay rates we can write the dressed effective operators as
\begin{eqnarray}
\hat{L}^{\gamma,0,\{1,2\}}_{\rm eff} &= &\pm i \sqrt{\gamma_{d}} \ket{T}\bra{T} + i \sqrt{\gamma_{d}} \ket{S}\bra{T}\mp\\&~ &\mp\tilde{\chi}_{a}\ket{T}\bra{00} - \tilde{\chi}_{a}\ket{S}\bra{00}\mp\tilde{\chi}_{a}^{*}\ket{T}\bra{11}
 - \tilde{\chi}_{a}^{*}\ket{S}\bra{11}\\
\hat{L}^{\gamma,1,\{1,2\}}_{\rm eff} &= &\mp\sqrt{2}\tilde{\chi}_{a}\ket{11}\bra{00}\mp\sqrt{2}\tilde{\chi}_{a}^{*}\ket{11}\bra{11} \pm i \sqrt{2 \gamma_{d}} \ket{11}\bra{T}\\
\hat{L}^\kappa_{\rm eff} &= &+\sqrt{\kappa_{\rm eff}} \ket{11}\bra{S} - 2 \sqrt{\kappa_{\rm eff}} \ket{S}\bra{00}
\end{eqnarray}
with the previous but shifted effective spontaneous emission rate $\gamma_{d}=\frac{\Omega^2}{8 \gamma} \frac{\gamma ^2+2 \Omega_{\rm MW}^2}{\gamma ^2+6 \Omega_{\rm MW}^2}=\gamma_{\rm eff} \cdot \eta$ and an additional spontaneous emission process activated by $\Omega_{\rm MW}$, with an amplitude $\tilde{\chi}_{\rm a}=\frac{\Omega \Omega_{\rm MW}}{2\sqrt{\gamma}}\frac{\gamma -i \sqrt{2} \Omega_{\rm MW}}{\gamma ^2+6 \Omega_{\rm MW}^2}$. Here, $\tilde{\chi}_{\rm a}$ carries a phase; the according effective decay rate is defined by $\gamma_{\rm a}=|\tilde{\chi}_{\rm a}|^2$, with $\gamma_{\rm a} \ll \gamma_{\rm d}$.
The effective cavity decay $\kappa_{\rm eff} = \frac{\kappa\Omega^2}{8 g^2}$ as the main loss process remains unaffected by the dressing since $\Omega_{\rm MW} \ll g$. The shifts of the effective Hamiltonian are again negligible so that $\hat{H}_{\rm eff} \approx \hat{H}_{\rm g}$.
\begin{figure*}[ht]
\centering
\includegraphics[width=12cm]{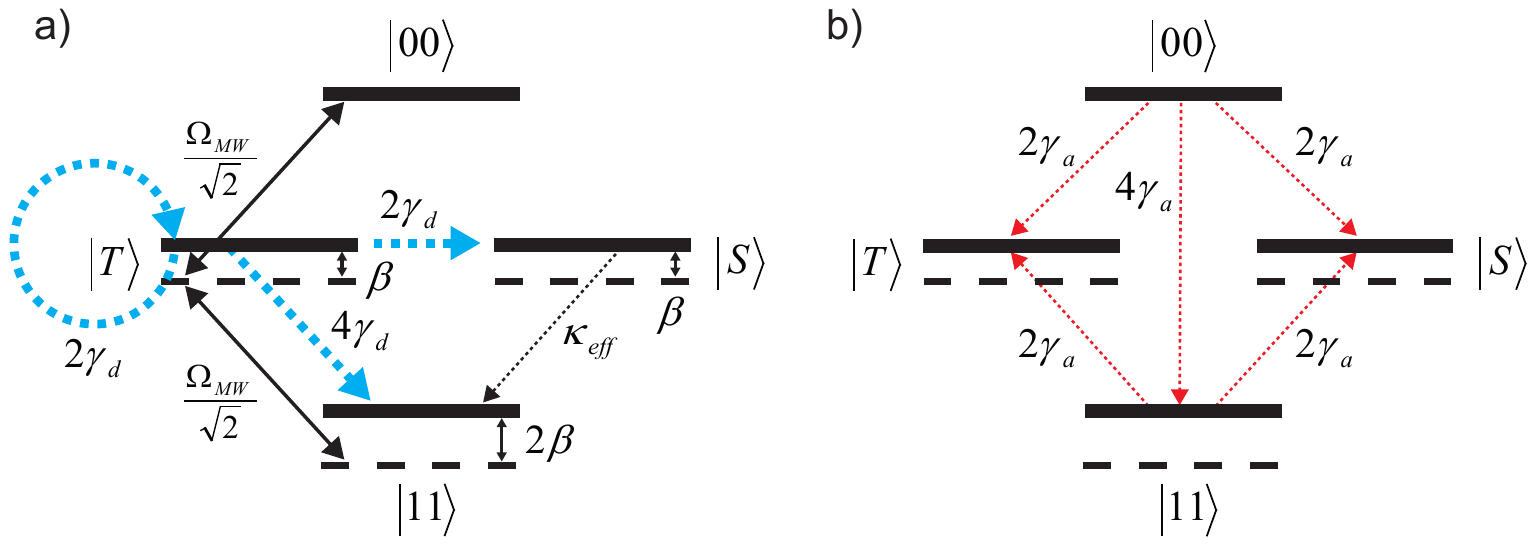}
\caption{Effective processes at increased driving. (a) Unaffected (black) and reduced (blue) processes. A detuning $\beta = \Omega_{\rm MW}/\sqrt{2}$ and dephasing $2 \gamma_{d}$ of state $\ket{T}$ retard the recycling of population $\ket{11} \rightarrow \ket{T} \rightarrow \ket{S}$. (b) Effective decay processes activated by dressing of the ground states by $\Omega_{\rm MW}$ (red).}
\label{FigStrong}
\end{figure*}\\
The effects of increased driving are illustrated in Fig. \ref{FigStrong}. The dressing of the triplet ground states $\ket{T_\pm}$ by $\Omega_{\rm MW}$ causes a reduction of the advantageous decay processes to $\gamma_{d}$ from $\gamma_{\rm eff}$, while the amplitude of the detrimental process $\kappa_{\rm eff}$ remains unchanged. The scaling of $\gamma_{\rm d}$ with $\eta$ is the result of averaging over the decay rates of the dressed triplet states $\gamma^{(T_l)}_{\rm eff}$ through back-transform. In addition, new decay channels at rates of $\gamma_{\rm a}$ are activated for high $\Omega_{\rm MW}$ and increase the error of the protocol by accumulation of population in state $\ket{11}$.

\subsection*{Derivation of the recycling error and optimal reshuffling}
\label{Strong}
Above we have derived the static error and the driving-dependent error originating from the shift of the ground states. An additional error emerges from the dynamics of the coherences which is not caught by rate equations of the populations:\\
From Fig. \ref{FigS1Evolution} a), c) we can see that accumulation of population in $\ket{11}$ is the bottleneck of the scheme. Coherent reshuffling $\Omega_{\rm MW}$ from $\ket{11}$ and $\ket{T}$ is used to recycle the population of $\ket{11}$. Hence, the additional error of accumulation of $\ket{11}$ is more pronounced the weaker $\Omega_{\rm MW}$ is compared to $\Omega$, regardless of the driving regime. The effective dephasing of $\ket{T}$ at a rate of $\gamma_{\rm d}$ and the detuning of $\beta=\Omega_{\rm MW}/\sqrt{2}$, however, tend to slow down the recycling process. Still, justified by its rapid decay of $6 \gamma_{d}$ altogether, the state $\ket{T}$ can be considered transient and can hence be adiabatically eliminated. In addition, we can ignore the evolution of $\ket{00}$ and the much weaker effective decay processes activated by $\Omega_{\rm MW}$ which have rates $\gamma_{\rm a}$. After adiabatic elimination of the rapidly dephasing coherences ($\dot{\rho}_{11,T}, \dot{\rho}_{T, 11}\approx 0$) and the population of $\ket{T}$ ($\rho_{T,T} \ll \rho_{11,11}$) we can write the dynamics in terms of two rates affecting the population of the desired singlet state
\begin{eqnarray}
\dot{P_S} \approx -\kappa_{\rm eff} P_S + \frac{8 \gamma_{\rm d} \Omega_{\rm MW}^2}{96 \gamma_{d}^2} P_{11}.
\end{eqnarray}
Thus, for the steady state ($\dot{P_S}=0$, $P_S \approx 1$) we derive the error
\begin{eqnarray}
\label{EqRecyclingError}
\left(1-F_S\right)_{\rm recy} & \approx \frac{12 \kappa_{\rm eff} \gamma_{d}}{\Omega_{\rm MW}^2}.
\end{eqnarray}
In order to make sure that the errors of Eqs. (\ref{EqDressingError}) and (\ref{EqRecyclingError}) are actually sufficient to describe the fidelity of the protocol at increased driving we also derive the steady state from the Liouvillian dynamics. To this end we solve the master equation represented by the effective Liouvillian $\mathcal{L}_{\rm eff}$ for $\dot{\rho}_{j,k}=0$ for all $j$, $k$.\\
The extended decay rates $\gamma_{\rm d}$ and $\gamma_{\rm a}$ also hold for stronger driving. Since the shuffling $\Omega_{\rm MW}$ is still much lower than the spontaneous emission $\gamma$, we also neglect dephasing originating from the $\Omega_{\rm MW}$-activated processes ($\gamma_{\rm a}$), as well as dephasing at a rate of $\kappa_{\rm eff}$, in the presence of dephasing at rates of $\gamma_{\rm d}$ ($\gamma_{\rm a}, \kappa_{\rm eff} \ll \gamma_{\rm d}$). Normalizing the obtained expression for the steady state and expanding it up to the second order in $\Omega$ and $\Omega_{\rm MW}$, we extract the complete driving-dependent error as
\begin{eqnarray}
\label{EqError}
\left(1-F_S\right)_{\rm comb} &\approx \frac{3 \gamma  \kappa }{2 g^2}+\frac{6 \kappa  \Omega_{\rm MW}^2}{g^2 \gamma}+\frac{3 \kappa \Omega^4}{16 g^2 \gamma \Omega_{\rm MW}^2}
\\
&\equiv \left(1-F_S\right)_{\rm stat}+\left(1-F_S\right)_{\rm dres}+\left(1-F_S\right)_{\rm recy}.
\end{eqnarray}
This is exactly the sum of the driving-dependent errors of Eqs. (\ref{EqDressingError}) and (\ref{EqRecyclingError}), expanded for small $\Omega_{\rm MW}$. We see that in fact the errors decouple. As one of these terms scales as $\Omega_{\rm MW}^{+2}$ and the other as $\Omega_{\rm MW}^{-2}$, the optimum for $\Omega_{\rm MW}$ is a trade-off between fast recycling requiring large $\Omega_{\rm MW}$ and the need not to shift states out of resonance favoring small $\Omega_{\rm MW}$. We use the result of Eq. (\ref{EqError}) to derive an optimal reshuffling of
\begin{equation}
\label{EqOptO2}
\Omega_{\rm MW, opt}=\frac{\Omega}{2^{5/4}}.
\end{equation}
Inserting $\Omega_{\rm MW,opt}$ into Eq. (\ref{EqError}) we obtain the combined error
\begin{eqnarray}
\label{EqDrivingError2}
\left(1-F_S\right)_{\rm comb} &=\frac{3}{2C}\left(1+\sqrt{2}\left(\frac{\Omega}{\gamma}\right)^2\right).
\end{eqnarray}
We will use this result below to discuss the scaling of the error with the speed of convergence and with the preparation time.\\
In Fig. \ref{FigDriving} a) we plot the analytical result for $\left(1-F_S\right)_{\rm comb}$ together with numerical curves obtained by extracting the steady state from the full and the effective Liouvillian for different optical driving $\Omega$, using the parameters of Ref. \cite{Kubanek1}, $(\gamma, \kappa)=(\frac{3g}{8},\frac{5g}{32})$, corresponding to a cooperativity of $C \approx 17$. Note that for the analytical curve in Fig. \ref{FigDriving} a) we have not discarded terms of higher order in $C^{-1}$, as in Eq. (\ref{EqError}), but have kept terms up to second order in $C^{-1}$ after solving for the steady state of the Liouvillian. For higher cooperativies, the higher orders become negligible and the expression for $\left(1-F_S\right)_{\rm comb}$ reduces to Eq. (\ref{EqError}).\\
Fig. \ref{FigDriving} b) contains the analytical and numerical results for the spectral gap, as a measure for the convergence rate. We find that for our initial assumption of weak driving ($\Omega \ll \gamma$) the analytic results for the scaling of both important performance measures, error and spectral gap, with the driving strength are very accurate. In addition, we find very good agreement with numerical results obtained from both the full and the effective master equation even up to an increased driving of $\Omega \approx \gamma/2 \approx 0.2g$. Despite the increased driving, the population of the excited states, in particular the close-to-resonant $\ket{S_1}$, does not exceed $\sim 5 \%$ for $\Omega = \gamma/2$ (cavity parameters of Ref. \cite{Kubanek1}) so that both the initial truncation of the Hilbert space to ground states and singly excited states as well as the concept of the effective dynamics of the ground states are well-justified even in the regime of increased driving.

\subsection*{Performance of the scheme at increased driving}
\begin{figure*}[ht]
\centering
\includegraphics[width=15cm]{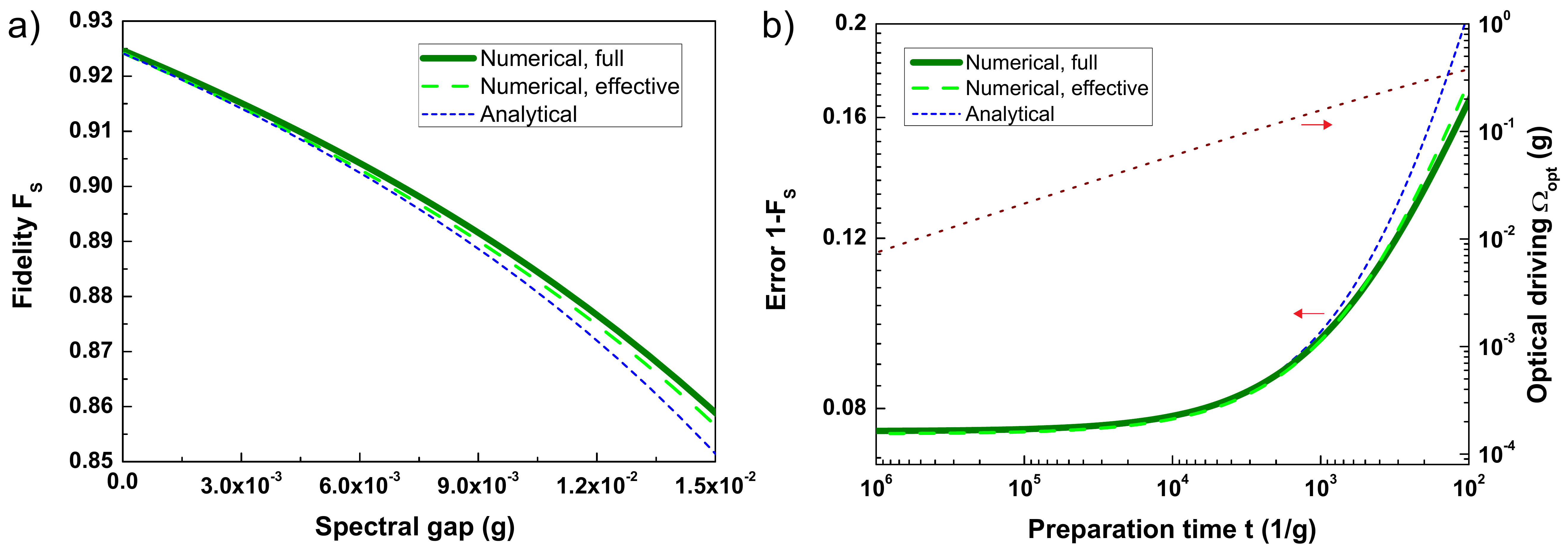}
\caption{Performance of the dissipative state preparation at increased driving. a) The fidelity of the steady state is lowered by the increase of the dynamic error when the spectral gap increases. (b) Error of the protocol (left axis) and optimal driving strength (right axis) vs. desired preparation time. Analytic results (blue, short dash) are in good agreement with numerical curves obtained from the full (dark green) and effective Liouvillian (green dash). The cavity parameters $(\gamma, \kappa)=(\frac{3g}{8},\frac{5g}{32})$ \cite{Kubanek1} ($C \approx 17$) and the optimized driving parameters $\Omega=2^{5/4}\Omega_{\rm MW}=2^{7/4}\beta$ were used; in (b) we also use and plot the optimized optical driving $\Omega_{\rm opt}$ (red dot) of Eq. (\ref{EqOmegaOpt}).}
\label{FigPerformance}
\end{figure*}

We evaluate the performance of the scheme, this time at increased driving, by estimating the trade-off between fidelity and convergence time. To this end, we use the results for the driving-dependent spectral gap and error of Eqs. (\ref{EqDrivingGap}) and (\ref{EqDrivingError2}) from which we eliminate the driving $\Omega$. In doing so we find
\begin{eqnarray}
\left(1-F_S\right)_{\rm per} &\approx \frac{3}{2 C} \frac{2 \sqrt{2} \gamma + 21 \lambda}{2 \sqrt{2} \gamma - 27 \lambda}&\overset{\lambda \ll \gamma}{\approx} \frac{3}{2C} + \frac{18 \sqrt{2} \kappa  \lambda }{g^2}.
\end{eqnarray}
For strong coupling $g \gg \left(\gamma,\kappa\right)$, or sufficiently high cooperativities $C \gg 10$, the static and the dynamic error decouple when the expression is expanded in $\lambda$. Thereby we obtain the slope of the tangent of $F_S$ for a small spectral gap $\lambda$. The analytic result shown in Fig. \ref{FigPerformance} agrees very well with the numerical results obtained from the full and dressed effective master equation as long as the assumption of perturbative optical driving is justified. For very rapid state preparation, the analytic expressions reproduce the decreasing trend of the numerical curves correctly. In Sec. \ref{SectionComparison} we compare these benchmarks for the scheme at hand with the ones for the schemes presented in Sec. \ref{SectionT0}, Sec. \ref{AppWS} and Sec. \ref{AppS0}.

\subsection*{Scaling of the dynamic error with the preparation time}
In the discussed setting, the scaling of the error and spectral gap provides an estimate of how fast the population decays into a desired steady state and to which extent the fidelity is lowered by an increased driving. For preparation within a fixed time one will thus have to make a compromise between convergence rate and the detrimental effects of increased driving. These two effects can be used to derive the optimal driving for a desired preparation time. 
To this end, the error of the protocol with respect to the preparation time $t$, consisting of a static and a dynamic part, can be written as
\begin{eqnarray}
\label{EqErrorTime}
\left(1-F_S\right)\left(\Omega, t\right)=\frac{3}{2C} + f \Omega^2 + \frac{3}{4} e^{-\Omega^2 t / r}.
\end{eqnarray}
with $f$ and $r$ specified below. Here we have assumed that the evolution begins in a complete statistical mixture of the four states. We minimize the error for a given state preparation time $t$ by taking its derivative with respect to $\Omega^2$
and obtain for the optimal driving strength
\begin{eqnarray}
\label{EqOmegaOpt}
\Omega_{\rm opt}^2 &=- \frac{r}{t} \log \frac{4fr}{3t}.
\end{eqnarray}
Above we have derived the combined static and dynamic error and the spectral gap (Eqs. (\ref{EqDrivingError2}) and (\ref{EqDrivingGap})) which we now associate with the quantities $f$ and $r$. With $f=\frac{3 \kappa}{\sqrt{2} g^2 \gamma}$ and $r = 12 \gamma$ we get $fr=\frac{36 \kappa}{\sqrt{2} g^2}$.
We thus obtain for the optimized error of the protocol
\begin{eqnarray}
\left(1-F_S\right)_{\rm opt}(t)&=\frac{3}{2C} + \frac{36 \kappa}{\sqrt{2} g^2 t} \left(1+ \log \frac{\sqrt{2} g^2 t}{48 \kappa}\right) \nonumber\\
&= \frac{3}{2C} + \frac{36}{\sqrt{2}}\sqrt{\frac{\kappa}{\gamma}} \sqrt{\frac{1}{C}}\frac{1}{g t} \left(1 + \log \left(\frac{\sqrt{2}}{48}\sqrt{\frac{\gamma}{\kappa}} \sqrt{C} g t\right)\right).
\end{eqnarray}
In the second line we have expressed the optimized error in terms of the cooperativity and the ratio of the decay rates $\gamma$ and $\kappa$. We find that -- apart from the linear static error ($\frac{3}{2C}$) -- the above expression for the optimized error also exhibits a favorable scaling of the dynamic error part with the square-root of the cooperativity. We plot this analytical result in Fig. \ref{FigPerformance} b), together with curves obtained numerically from the full/effective Liouvillian, using the cavity parameters $(\gamma, \kappa)=(\frac{3g}{8},\frac{5g}{32})$ \cite{Kubanek1} ($C \approx 17$) and the optimized driving parameters $\Omega=2^{5/4}\Omega_{\rm MW}=2^{7/4}\beta$, as well as the optimized optical driving $\Omega_{\rm opt}$ of Eq. (\ref{EqOmegaOpt}). We find good agreement even for reasonably short preparation times $\approx 10^3g^{-1}$ where we get fidelities above $90 \%$.

\section{Schemes for various experimental situations}
\label{SectionT0}
The $\ket{S_1}$ scheme for effective spontaneous emission mediated by a dark state, discussed in the preceding sections, as well as the $\ket{S_0}$ scheme, discussed in Sec. \ref{AppS0}, both assume a phase difference of $\phi=\pi$ between the optical driving of the two atoms. While in present-day cavity experiments the position of the atoms along the cavity axis with respect to the cavity standing wave is well-controllable within the Lamb-Dicke regime, their transversal position is not necessarily confined. If the atoms are driven by laser fields oriented transverse to the cavity axis this results in a random phase factor $e^{i k \cdot r(t)}$ (with wave vector $k$ and relative position of the atoms $r(t)$).
Hence, the assumption of a relative and stable phase relation $\phi$ rules out common transversal driving of the atoms in the absence of transversal trapping.\\
This section deals with better suited alternatives for today's cavity experiments: We present a $\ket{T_0}$ and a $\ket{T_1}$ scheme that can be implemented with a driving with $\phi=0$.\\
\\
This phase relation can be obtained by driving the cavity with a strong laser which is strongly detuned from a cavity mode but near resonance with the atomic transition $|0\rangle \leftrightarrow |e \rangle$. The detuned drive creates a coupling mediated by the off-resonant cavity mode and the phase relation will then be set by the phase of the cavity mode. If the cavity driving mode and the mode used to create the entanglement are commensurate, this will ensure that we have the phase relation $\phi=0$.\\
\\
As an alternative, we also discuss the possibility to use common addressing by a transverse laser at an arbitrary relative phase $\phi$, which results in an effective combination of the $\ket{T_0}$ and $\ket{S_0}$ scheme.

\subsection*{A scheme for engineered decay mediated by $\ket{T_0}$}
For the scheme at hand, we choose to use the subspace containing the atomic excited state $\ket{T_0}$ and the cavity-excited state $\ket{T}\ket{1}$, in order to realize strong spontaneous emission from $\ket{00}$ into $\ket{S}$, mediated by $\ket{T_0}$. For this $\ket{T_0}$ scheme we use non-vanishing laser ($\Delta$) and cavity ($\delta$) detuning, but a vanishing detuning $\beta=0$ of the microwave/Raman field. This means that $\tilde{\Delta}=\tilde{\Delta}_n$ and $\tilde{\delta}=\tilde{\delta}_n$ for all $n$.

\begin{figure*}[ht]
\centering
\includegraphics[width=14cm]{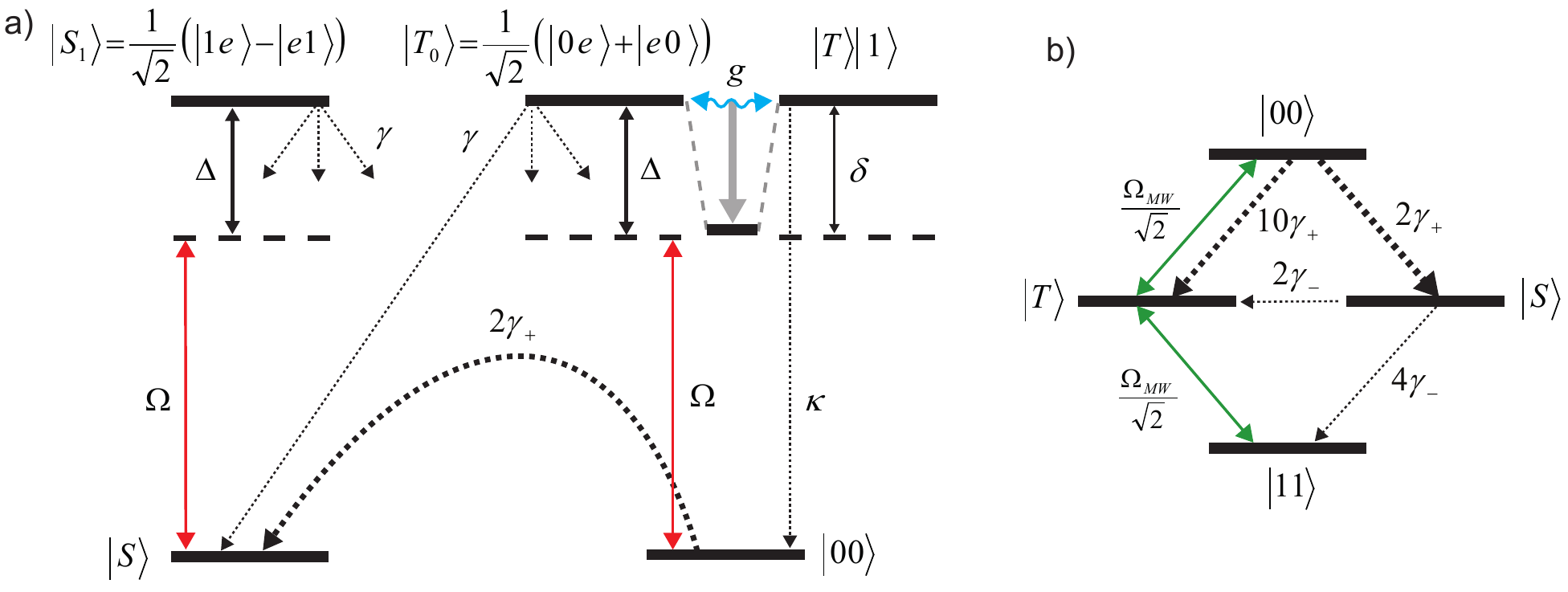}
\caption{Mechanism and effective processes of the $\ket{T_0}$ scheme. (a) Dissipative preparation of the maximally entangled singlet state $\ket{S}$. For an appropriate choice of atomic and cavity detuning $\Delta$ and $\delta$ the cavity interaction of strength $\sqrt{2} g$ enhances the effective spontaneous emission $\gamma_{\rm eff}$ from state $\ket{00}$ by shifting the lower dressed state of $\ket{T_0}$ and $\ket{T}\ket{1}$ into resonance with the optical driving $\Omega$. States $\ket{T}$ and $\ket{11}$ are coupled to $\ket{00}$ by a microwave field or Raman transition $\Omega_{\rm MW}$ (not shown). Effective decay from $\ket{S}$ is suppressed as it involves the atom-cavity interaction dark state $\ket{S_1}$, the detuning $\Delta$ of which is not compensated. (b) Effective level scheme and ground state to ground state processes for the presented scheme.}
\label{FigMechanismT0}
\end{figure*}

\subsubsection*{Mechanism of the state preparation}
The working principle is illustrated in Fig. \ref{FigMechanismT0}.  Population from state $\ket{00}$ is excited to $\ket{T_0}$ by a weak optical field of strength $\Omega$. The atomic excited state $\ket{T_0}$ is coupled to the cavity excited state $\ket{T}\ket{1}$ by the atom-cavity interaction $\hat{H}_{\rm ac}$. Due to the strong coupling ($g$), the states $\ket{T_0}$ and $\ket{T}\ket{1}$, initially detuned by $\tilde{\Delta}$ and $\tilde{\delta}$, form dressed states. Treating the detunings of the excited states as complex, as discussed in Sec. \ref{SectionMethods}, the energies of these dressed states can be written as
\begin{eqnarray}
\tilde{E}_\pm=\frac{\tilde{\Delta}+\tilde{\delta}}{2}\pm\frac{1}{2}\sqrt{\left(\tilde{\Delta}+\tilde{\delta}\right)^2-4\left(\tilde{\Delta}\tilde{\delta}-g^2\right)},
\end{eqnarray}
where $\tilde{\Delta} = \Delta-\frac{i\gamma}{2}$ and $\tilde{\delta} = \delta-\frac{i\kappa}{2}$.
We engineer an efficient spontaneous emission process that prepares the singlet state $\ket{S}$ by setting the cavity detuning equal to the cavity line shift $\delta = g^2/\Delta$. With this choice the lower dressed state of $\ket{T_0}$ and $\ket{T}\ket{1}$ is shifted close to resonance, ${\rm Re} (\tilde{E}_-)\approx 0$. Consequently, population from $\ket{00}$ -- which is coupled to $\ket{T_0}$ -- is rapidly transferred to $\ket{S}$ by spontaneous emission via the lower dressed state of $\ket{T_0}$ and $\ket{T}\ket{1}$. On the other hand, decay out of $\ket{S}$ involves excitation of $\ket{S_1}$. Since $\ket{S_1}$ is the only dark state of the atom-cavity interaction, its detuning $\Delta$ means that it is not in resonance. Hence, the decay into the singlet state $\ket{S}$ is engineered to be much stronger than the decay out of $\ket{S}$ so that the maximally entangled state $\ket{S}$ is efficiently prepared. The atomic detuning provides a trade-off between virtual character of the excited dressed states on the one hand, and spontaneous and cavity decay on the other hand; by setting $\Delta=g\sqrt{\frac{\gamma}{\kappa}}$ we minimize the decay width of the dressed states, $\tilde{E}_- \approx {\rm Im}(\tilde{E}_-) \approx \frac{i}{2} \left(\Delta\kappa+\delta\gamma\right)$. Furthermore, coherent coupling of the triplet states $\ket{00}$, $\ket{11}$, and $\ket{T}$ by the microwave/Raman field $\Omega_{\rm MW}$ guarantees that all triplet states decay rapidly towards the singlet state $\ket{S}$. 

\subsubsection*{Effective processes}
For the discussed scheme we have assumed the relative phase of the optical driving between the atoms to be zero ($\phi=0$).
Using this, the effective processes can be read off from Figs. \ref{FigInteraction} and \ref{FigPropagators}. Given our choice of the parameters $\delta$, $\Delta$ and $\beta$ the terms of the effective detunings can be simplified to obtain the scaling of the effective decay processes. Using $\tilde{g}_{1, \rm eff} \approx i \sqrt{\gamma\kappa}$, $\tilde{g}_{2, \rm eff} \approx g/\sqrt{2}$, $\tilde{\Delta}_{0, \rm eff} \approx \Delta$, $\tilde{\Delta}_{1, \rm eff} \approx - i \gamma$ and $\tilde{\Delta}_{2, \rm eff} \approx - \Delta$ the effective operators simplify to
\begin{eqnarray}
\hat{L}_{\rm eff}^{\gamma,0,\{1,2\}} &= i\sqrt{2\gamma_+} \ket{00}\bra{00} \pm \sqrt{\gamma_-} \ket{T}\bra{S} \\
\hat{L}_{\rm eff}^{\gamma,1,\{1,2\}} &= i \sqrt{\gamma_+} \left(\ket{T}\bra{00} \mp \ket{S}\bra{00}\right) \pm \sqrt{2 \gamma_-} \ket{11} \bra{S}\\
\hat{L}_{\rm eff}^\kappa &= - i \sqrt{\kappa_{\rm eff}} \ket{T}\bra{00}.
\end{eqnarray}
Here, the spontaneous emission processes $\hat{L}_{\rm eff}^{\gamma,0,\{1,2\}}$ transfer population from $\ket{00}$ into the desired state $\ket{S}$ at a strongly enhanced rate of $2\gamma_+=\frac{\Omega^2}{16 \gamma}$. Loss from the singlet state also occurs by spontaneous emission at a much weaker rate of $\gamma_-=\frac{\kappa \Omega^2}{32 g^2}$. As opposed to the $\ket{S_0}$ and $\ket{S_1}$ schemes, the effective cavity decay, here with a rate of $\kappa_{\rm eff}=\frac{\Omega^2}{4 \gamma}$, does not directly affect the singlet state. The effective processes are illustrated in Fig. \ref{FigMechanismT0} b).

\subsubsection*{Parameter and performance analysis}
Setting up the rate equations in the same manner as for the previous scheme is straightforward. We obtain for the error and spectral gap
\begin{eqnarray}
\left(1-F_S\right)_{\ket{T_0}} &= \frac{11}{2 C} \\
\lambda_{\ket{T_0}} &= \frac{2-\sqrt{3}}{8} \frac{\Omega^2}{\gamma}.
\end{eqnarray}
Both the error and the spectral gap are found to have the same scaling with the parameters of the system as the $\ket{S_1}$ scheme, but exhibit different proportionality factors. The performance of the schemes presented in this section is optimal for an $\Omega_{\rm MW}$ in the interval of $\Omega_{\rm MW}=\frac{\Omega}{2}$ to $\Omega_{\rm MW}=\frac{\Omega}{3}$; the latter value is used for the simulations below.

\subsection*{Laser driving with random relative phase between the atoms: the $\ket{T_0} / \ket{S_0}$ scheme}
One can conceive of experimental situations for which neither transversal confinement of the atoms (and hence a stable phase relation), nor cavity driving are available. In the following we argue that using laser addressing of the two atoms at random relative phase can be suitable for the preparation of an entangled steady state of high fidelity.\\
Apart from the driving phase $\phi$, the conditions for the operation of the $\ket{T_0}$ scheme of the previous section and the $\ket{S_0}$ scheme presented in Ref. \cite{KRS} which is briefly discussed in Sec. \ref{AppS0} are identical; in particular $\delta=g^2/\Delta$. Transversal drift of the atoms results in a random $\phi(t)$ with fluctuations much slower than the couplings of the system. Depending on the actual value of $\phi$ the driving either crosses the singlet/triplet subspace, $\ket{00} \overset{\Omega}{\rightarrow} \ket{S_0}$, as illustrated in Fig. \ref{FigS0Mechanism} or stays within the subspace, $\ket{00} \overset{\Omega}{\rightarrow} \ket{T_0}$, similar to Fig. \ref{FigMechanismT0}. Therefore, effective decay channels are instantaneously weighted with $\phi$, as $\gamma_{\rm eff}(t) = (\gamma_{\rm eff})_{\ket{T_0}} \cdot \cos^2 \phi(t)+(\gamma_{\rm eff})_{\ket{S_0}} \cdot \sin^2 \phi(t)$. Thus, the system mechanisms are an combination of the two individual schemes. Averaging the decay rates results in a combined error and spectral gap
\begin{eqnarray}
\left(1-F_S\right)_{\ket{T_0}/\ket{S_0}} &= \frac{9}{2 C} \\
\lambda_{\ket{T_0}/\ket{S_0}} &= \frac{(9-2\sqrt{3}-\sqrt{5})\Omega^2}{32 \gamma} \approx \frac{\Omega^2}{10 \gamma}.
\end{eqnarray}
We conclude that a setup with arbitrary driving phase is also suitable for an experimental realization of a high-fidelity entangled state of two atoms in an optical cavity.

\subsection*{Cavity driving: A scheme for engineered decay mediated by $\ket{T_1}$}
\label{SectionT0}
In the following we briefly discuss a possible $\ket{T_1}$ scheme that combines elements of the $\ket{T_0}$ and of the $\ket{S_1}$ scheme. It exhibits an improved error and spectral gap compared to the $\ket{T_0}$ scheme and is also suitable for cavity driving $\phi=0$, but not for transversal laser driving without transversal confinement.\\
Setting $\delta=\frac{2g^2}{\Delta}$ and accordingly, $\Delta=g \sqrt{\frac{2\gamma}{\kappa}}$, shifts one of the dressed states of $\ket{T_1}$ and $\ket{11}\ket{1}$ into resonance. Then, $\ket{S}$ is effectively prepared through $\ket{T_1}$ by spontaneous emission. Similar to the $\ket{S_1}$-scheme, a choice of $\beta=\frac{\Omega_{\rm MW}}{\sqrt{2}}$ guarantees an equal share of $\ket{T}$ in the dressed triplet states so that these states decay equally rapidly into $\ket{S}$.\\
As compared to the $\ket{T_0}$ scheme the contrast between the unwanted $\ket{S_1}$-mediated terms and the desired $\ket{T_1}$-mediated terms is more pronounced than previously for the decay through $\ket{S_1}$ and $\ket{T_0}$. The error and spectral gap are therefore improved compared to the $\ket{T_0}$ scheme:
\begin{eqnarray}
\left(1-F_S\right)_{\ket{T_1}} &= \frac{9}{2 C} \\
\lambda_{\ket{T_1}} &= \frac{\Omega^2}{48 \gamma}
\end{eqnarray}
The performance of this $\ket{T_1}$ scheme at increased optical driving will also be addressed numerically in the section below.

\section{Comparison of the presented schemes}
\label{SectionComparison}
In the following we provide an overview of the presented schemes and compare their error in the preparation of the maximally-entangled singlet state and their spectral gap as a measure for the rate of convergence. We separately discuss the scaling of the static error due to the imperfections of the cavity (as discussed in Sec. \ref{Weak}), and the dynamic error originating from increased optical driving by dressing of the levels (as in Sec. \ref{Strong}). An overview of all the schemes and in which section they can be found is shown in Table \ref{TableComparison} along with a few key results on the performance of each scheme.
\begin{figure*}[ht]
\centering
\includegraphics[width=15cm]{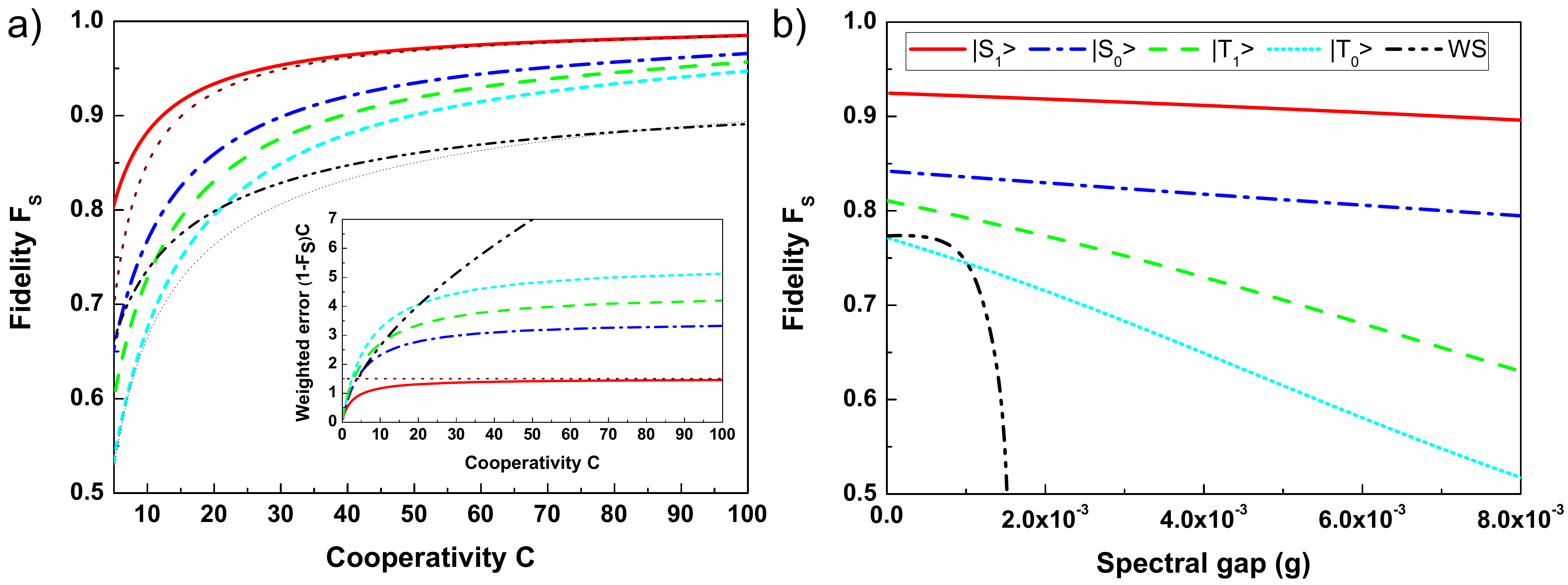}
\caption{Comparison of static and dynamic error for the presented schemes, obtained numerically from the full Liouvillian of Eq. (\ref{EqFullMaster}). (a) Scaling of the fidelity with the cooperativity (inset: error weighted with the cooperativity). The lowest error is found for the $\ket{S_1}$ scheme ($\frac{3}{2C}$, red solid), followed by the $\ket{S_0}$ scheme ($\frac{7}{2C}$, blue dash-dot) and the schemes suitable for cavity driving, $\ket{T_1}$ ($\frac{9}{2C}$, green dash) and $\ket{T_0}$ ($\frac{11}{2C}$, sky-blue short dash). Noticeable is the qualitative difference between linear scaling of these schemes and the square-root scaling law for the WS scheme ($\frac{3}{2\sqrt{2C}}$, black dash-dot-dot). Analytical results for the asymptotic scaling are shown for the $\ket{S_1}$ and WS scheme (red and black dots) (b) Fidelity vs. speed of convergence (spectral gap). A compromise between fidelity and speed limits the performance of the WS scheme, while close-to-linear scaling of both the $\ket{S_{1/0}}$ and $\ket{T_{1/0}}$ schemes allows rapid state preparation. For b) the cavity parameters $(\gamma, \kappa)=(\frac{3g}{8},\frac{5g}{32})$ \cite{Kubanek1}, with $C \approx 17$ were used, in a) $C$ was varied keeping the ratio $\gamma/\kappa=12/5$ constant. The optimized parameters used for each of the presented schemes are specified in the corresponding section. The same line format is used to denote the schemes in (a) and (b).}
\label{FigComparison}
\end{figure*}

\subsection*{Static error scaling with the cavity parameters}
In Fig. \ref{FigComparison} a) we have plotted the fidelity of the steady state with the maximally-entangled singlet state, as a function of the cooperativity $C=\frac{g^2}{\gamma \kappa}$ for all schemes presented in this work. The curves were obtained numerically by extracting the steady state from the full Liouvillian of Eq. (\ref{EqFullMaster}), using the optimized parameters specified in the corresponding section. For the error scaling of the $\ket{S_1}$ and the WS scheme, we plot the analytical curves along with the numerical ones. The linear scaling of the static error with $C$ is more clearly seen in the inset, where we plot the weighted error $\left(1-F_S\right)C$. In agreement with our analytic results, we find this quantity to be independent of $C$ for $C \gg 10$ for the $\ket{T_0}$, $\ket{T_1}$, $\ket{S_0}$ and $\ket{S_1}$ schemes, while the adapted Wang-Schirmer (WS) scheme of App. \ref{AppWS} exhibits an error scaling $\propto \sqrt{C}$ as is the case for coherent unitary protocols\footnote{See \cite{SM} for a detailed discussion of this point.}. The best error scaling of $\frac{3}{2C}$ is provided by the $\ket{S_1}$ scheme; an increase of the error to $\frac{7}{2C}$ is found for the $\ket{S_0}$ scheme. The schemes that are suitable for cavity driving in the absence of transversal confinement, $\ket{T_1}$ and $\ket{T_0}$ also exhibit linear scaling with the cooperativity with further increasing proportionality factors $\frac{9}{2C}$ and $\frac{11}{2C}$. Above cooperativities of $C \approx 10$ the square-root scaling error of the WS scheme $\frac{3}{2\sqrt{2C}}$ is outperformed by the $\ket{T_1}$ scheme which uses similar conditions. An overview and numerical examples are given in Table \ref{TableComparison}.

\subsection*{Dynamic error scaling with the speed of convergence}
In addition to the static error scaling of Fig. \ref{FigComparison} a) we present the dynamic error scaling with the spectral gap in Fig. \ref{FigComparison} b). These curves were obtained by numerically extracting the spectral gap from the full Liouvillian of Eq. (\ref{EqFullMaster}).\\
Again, the best performance is shown by the $\ket{S_1}$ scheme, followed by the other three schemes which all exhibit an almost linear scaling; the schemes suitable for cavity driving, $\ket{T_1}$ and $\ket{T_0}$, have a steeper slope. On the other hand, the performance of the adapted WS scheme is governed by a compromise between fidelity and speed, that also affects the preparation speed. Here, the fidelity with the entangled state drops considerably at increased speed, so that the state preparation of the WS scheme is found to be slow (cf. Ref. \cite{WS}). Numerical examples of the performance are also given in Table \ref{TableComparison}.\\
\\
We conclude that all relevant benchmarks, both static and dynamic are best for the $\ket{S_1}$ scheme that was discussed in detail in Sec. \ref{SectionS1} and \ref{SectionSpeed}. Theoretically, this scheme allows for the generation of a maximally-entangled state with fidelities of more than $90\%$ and convergence time of about 10 $\mu s$ for present-day cavity experiments.\\
Yet, for a possible experimental realization of steady-state entanglement in optical cavities in the absence of transversal confinement of the atoms, the cavity-driven schemes $\ket{T_1}$ and $\ket{T_0}$ and the randomly laser-driven $\ket{T_0}$/$\ket{S_0}$ scheme are more suitable. Despite the lower proportionality factors, these schemes provide preparation of an entangled state with convergence times of a few tens of microseconds at fidelities $\approx 80 \%$ with present day optical cavities.
\begin{table}
\caption{Comparison of the discussed schemes: Analytic scaling of the error and rate of convergence (spectral gap) with respect to the cavity parameters for $C \gg 10$; comparative numbers for a cavity QED system as in Ref. \cite{Kubanek1}, $(g,\gamma,\kappa)/2\pi=(16,6,2.5)$ MHz, $C \approx 17$. Characteristic dynamic measures are given for a driving that causes a dynamic error of $2 \%$.}
\footnotesize\rm
\begin{tabular*}{\textwidth}{@{}l*{15}{@{\extracolsep{0pt plus12pt}}l}}
\br
Scheme&Static error&Spectral gap&Convergence time&Transversal confi-\\
&max. fidelity&at $2 \%$ error&at $2 \%$ error&nement required?\\
\mr
$\ket{S_1}$ (Sec. \ref{SectionS1})&$3 / 2 \ C^{-1}$&$\Omega^2 / 12 \gamma$&&\\
&92.5$\%$&$6 \cdot 10^{-3}g$&$10 \ \mu s$&yes\\
$\ket{S_0}$ (Sec. \ref{AppS0})&$7 / 2 \ C^{-1}$&$(5-\sqrt{5}) \Omega^2 / 16 \gamma$&&\\
&84.2$\%$&$3 \cdot 10^{-3}g$&$20 \ \mu s$&yes\\
$\ket{T_1}$ (Sec. \ref{SectionT0})&$9 / 2 \ C^{-1}$&$\Omega^2 / 48 \gamma$&&\\
&81.1$\%$&$1 \cdot 10^{-3}g$&$60 \ \mu s$&no\\
$\ket{T_0}$ (Sec. \ref{SectionT0})&$11 / 2 \ C^{-1}$&$(2 - \sqrt{3})\Omega^2 / 8 \gamma$&&\\
&77.2$\%$&$8 \cdot 10^{-4}g$&$80 \ \mu s$&no\\
$\ket{T_0}/\ket{S_0}$ (Sec. \ref{SectionT0})&$9 / 2 \ C^{-1}$&$\Omega^2 / 10 \gamma$&&\\
&79.7$\%$&$1 \cdot 10^{-3}g$&$60 \ \mu s$&no\\
WS (Sec. \ref{AppWS})&$3 / 2 \ \sqrt{2C}^{-1}$&$2 g^2 \Omega^2 / 3 \Delta^2 \kappa$&&\\
&77.3$\%$&$9 \cdot 10^{-4}g$&$70 \ \mu s$&no\\
\br
\label{TableComparison}
\end{tabular*}
\end{table}

\section{Conclusion and outlook}
In this article, we have performed a detailed study of the dissipative preparation of a highly entangled steady state of two $\Lambda$-atoms in a single-mode optical cavity by engineering the naturally occurring sources of noise: spontaneous emission and cavity loss. We have employed an effective operator formalism to identify and understand the effective decay processes. The schemes we have proposed and analyzed use various engineered effective decay processes of either spontaneous emission or cavity loss to rapidly reach a maximally entangled singlet state as the steady state of the dissipative time evolution at high fidelity.\\
Our schemes are suitable for various experimental situations and require coherent driving by only a single laser field and another microwave or Raman field; in particular we have proposed schemes which work in the absence of trapping of the atoms in the cavity in the transversal direction, some of which are tailored for cavity driving.\\
We have shown that all our schemes provide a favorable scaling of the static error that is linear with the cooperativity of the cavity. In addition we derived the scaling of the dynamic error and resolved their underlying mechanisms. Building upon our results we have investigated the optimal conditions for the preparation of an entangled steady state for a given preparation time.\\
We consider our study relevant for the demonstration of an entangled steady state by means of dissipation in today's cavity QED experiments. A thorough theoretical understanding of the mechanisms allowing for dissipative state preparation of two qubits is important as a stepping stone for more complicated studies involving many particles.

\section*{Acknowledgements}
We thank the Rempe group and Eran Kot for helpful discussions. This work was supported by the Danish National Research Foundation, the Villum Kann Rasmussen Foundation and the European project QUEVADIS. FR acknowledges support from the DAAD (German Academic Exchange Service); MJK acknowledges support from the Niels Bohr International Academy.

\section[Appendix A: Wang-Schirmer scheme generalized to $\Lambda$-atoms]{Appendix A: Wang-Schirmer (WS) scheme generalized to $\Lambda$-atoms}
\label{AppWS}
\begin{figure*}[ht]
\centering
\includegraphics[width=12cm]{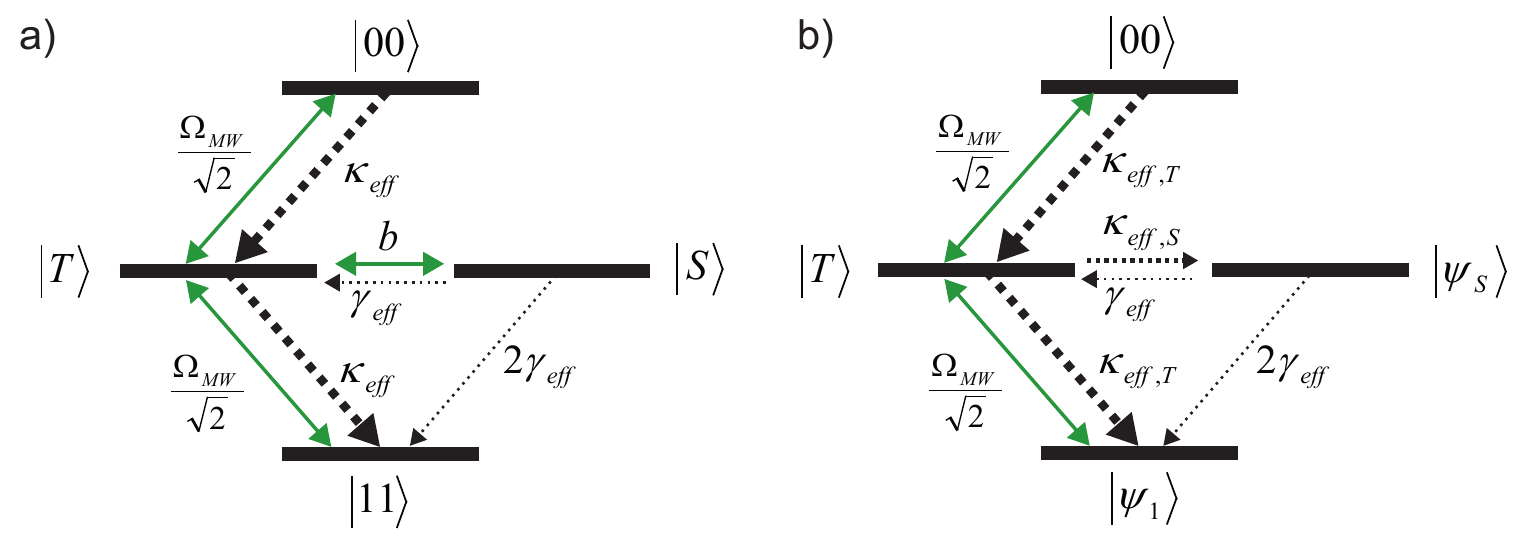}
\caption{Effective processes of the Wang-Schirmer scheme adapted to $\Lambda$-atoms. (a) In the shuffling picture, the triplet states are coupled by a microwave or Raman transition $\Omega_{\rm MW}$ and decay by an effective cavity decay $\kappa_{\rm eff}$. The singlet $\ket{S}$ is coherently coupled to $\ket{T}$ by the level shift $b$. (b) In the steady-state picture, the desired steady state $\ket{\psi_{S}}$ is no longer coherently coupled, but dissipatively prepared from $\ket{T}$ at a rate of $\kappa_{{\rm eff},S}$. Population in $\ket{\psi_S}$ is lost by spontaneous emission $\gamma_{\rm eff}$.}
\label{FigEffWS}
\end{figure*}

In the following we generalize the scheme of Wang and Schirmer \cite{WS}, originally proposed for two two-level atoms, to $\Lambda$-type atoms. In contrast to the schemes presented so far, the ground state $\ket{1}$ of the two atoms is shifted asymmetrically for the two atoms so that a coherent coupling is created between $\ket{S}$ and $\ket{T}$. This means that a pure singlet state $\ket{S}$ can, even in the absence of spontaneous emission, no longer be reached as the steady state of the time evolution. However, an engineered cavity decay process between the triplet states $\ket{00} \rightarrow \ket{T} \rightarrow \ket{11}$ is used to prepare a steady state which has a minor overlap with $\ket{11}$ and a high fidelity with the singlet.\\
By elimination of the excited states we will reduce the coupled $\Lambda$-atom systems to an effective system of two coupled qubits described by the master equation of Ref. \cite{WS} and, subsequently, derive the error scaling of the preparation of the entangled state with the cavity parameters.\\
The coherent interactions are given by the Hamiltonian of the system as in Eqs. (\ref{EqH1}-\ref{EqH2}). Here, we use a ground-state Hamiltonian $\hat{H}_{\rm g}$
\begin{eqnarray}
\hat{H}_{\rm g} &= \sum_{j=1,2} \left( \beta + (-1)^j b \right) \ket{1}_j\bra{1} + \Omega_{\rm MW} \left( \ket{0}_j\bra{1} + \ket{1}_j\bra{0} \right),
\end{eqnarray}
where a static magnetic field $b$ has been introduced that results in a shift of ground state $\ket{1}$ with opposite signs for the two atoms. The driving fields for both atoms exhibit the same phase ($\phi=0$) so that the general effective Lindblad operators are the same as for the $\ket{T}$ schemes. The effective Hamiltonian consists of shifts of the ground states
\begin{eqnarray}
\label{EqHWS}
\hat{H}_{\rm eff} = &- {\rm Re} \left[\frac{\Omega^2}{2\tilde{\Delta}_{1, \rm eff}}\right] \ket{00}\bra{00} - {\rm Re} \left[\frac{\Omega^2}{4 \tilde{\Delta}_{2, \rm eff}}\right]\ket{T}\bra{T} - \nonumber \\ &- {\rm Re} \left[\frac{\Omega^2}{4 \tilde{\Delta}_{0, \rm eff}}\right] \ket{S}\bra{S} + \hat{H}_{\rm g}.
\end{eqnarray}
A parameter choice of $\delta=0$, $\Delta \gg g \gg \left(\gamma, \kappa, \Omega, \Omega_{\rm MW}, \beta, b\right)$ and $\Delta \kappa \gg g^2$, allows for an adiabatic elimination of the excited atomic levels $\ket{e}$. In this limit, all propagators $\tilde{\Delta}_{n, \rm eff}^{-1}$ are simply determined by the shifts of the atomic excited levels $\Delta$,
\begin{eqnarray}
\tilde{\Delta}_{n, \rm eff}^{-1} &\approx \Delta^{-1}.
\end{eqnarray}
We then obtain the effective Lindblad operators
\begin{eqnarray}
\label{EqDicke}
\hat{L}_{\rm eff}^\kappa &= &-i \sqrt{\kappa_{\rm eff}} \ket{11}\bra{T} - i \sqrt{\kappa_{\rm eff}} \ket{T}\bra{00} \\
\hat{L}_{\rm eff}^{\gamma,0,\{1,2\}} &= &+ \sqrt{2\gamma_{\rm eff}} \ket{00}\bra{00} + \sqrt{\gamma_{\rm eff}/2} \left(\ket{T}\bra{T} \pm \ket{T}\bra{S}\right) + \nonumber \\ && + \sqrt{\gamma_{\rm eff}/2} \left(\pm \ket{S}\bra{T} + \ket{S}\bra{S} \right) \\
\hat{L}_{\rm eff}^{\gamma,1,\{1,2\}} &= &+ \sqrt{\gamma_{\rm eff}} \left(\mp\ket{S}\bra{00}+\ket{T}\bra{00}\right) \pm \sqrt{\gamma_{\rm eff}} \ket{11}\bra{S} + \nonumber \\
&& + \sqrt{\gamma_{\rm eff}} \ket{11}\bra{T},
\end{eqnarray}
where $\kappa_{\rm eff}=\frac{2 g^2 \Omega^2}{\Delta^2 \kappa}$ and $\gamma_{\rm eff}=\frac{\gamma \Omega^2}{16 \Delta^2}$.
The effective Hamiltonian is given by
\begin{eqnarray}
\label{EqHWS2}
\hat{H}_{\rm eff} = &-\frac{\Omega^2}{2 \Delta} \ket{00}\bra{00} + 2 \beta \ket{11}\bra{11} + \nonumber \\ &+ \left(\beta-\frac{\Omega^2}{4 \Delta}\right) \ket{T} \bra{T} + \left(\beta -\frac{\Omega^2}{4 \Delta}\right) \ket{S}\bra{S} + \nonumber\\
&-b \left( \ket{S}\bra{T} + \ket{T} \bra{S} \right) + \Omega_{\rm MW} \left(\ket{00}\bra{T} + \ket{T}\bra{11} + H.c.\right).
\end{eqnarray}
The corresponding effective couplings are shown in Fig. \ref{FigEffWS} a).\\
In order to match the master equation of Ref. \cite{WS}, we compensate the shifts in the effective Hamiltonian in Eq. (\ref{EqHWS2}); i.e. we set the (symmetric) detuning of $\ket{1}$ to $\beta=-\frac{\Omega^2}{4 \Delta}$. We then obtain the effective Hamiltonian and the effective cavity decay
\begin{eqnarray}
\hat{H}_{\rm eff}&=-b \left( \ket{S}\bra{T} + \ket{T} \bra{S} \right) + \Omega_{\rm MW} \left(\ket{00}\bra{T} + \ket{T}\bra{11} + H.c.\right) \\
\hat{L}^{\kappa}_{\rm eff}&=\sqrt{\kappa_{\rm eff}} \left(\ket{11}\bra{T} + \ket{T}\bra{00}\right).
\end{eqnarray}
From here we construct the Dicke-type master equation along the lines of Ref. \cite{WS}. In contrast to Ref. \cite{WS}, however, we analytically take the (for this scheme) undesired but unavoidable spontaneous emission into account and write
\begin{eqnarray}
\dot{\rho}=-i\left[\hat{H}_{\rm eff},\rho\right]+\mathcal{D}[\hat{L}^\kappa_{\rm eff}]+\sum_k \mathcal{D}[\hat{L}^{\gamma,k}_{\rm eff}]\label{HeffWS}
\end{eqnarray}
with $\hat{L}_\kappa$ as the engineered decay and $\hat{L}^{\gamma}_{k}$ as the undesirable spontaneous emission processes.
In order to analyze the scheme we note that 
\begin{eqnarray}
\ket{\psi_{S}}= \frac{1}{\sqrt{\Omega_{\rm MW}^2+b^2}} \left(b\ket{11}+\Omega_{\rm MW}\ket{S}\right)
\end{eqnarray}
is a steady state of the Hamiltonian of Eq. (\ref{EqDicke}) and also of the Liouvillian of Eq. (\ref{HeffWS}) in the absence of spontaneous emission. To understand the dissipative state preparation mechanism we change into a basis consisting of $\ket{\psi_S}$ and the orthogonal state
\begin{eqnarray}
\ket{\psi_1}=\frac{1}{\sqrt{\Omega_{\rm MW}^2+b^2}} \left(\Omega_{\rm MW}\ket{11}-b\ket{S}\right).
\end{eqnarray}
As can be seen from Fig. \ref{FigEffWS} b), the singlet-like steady state $\ket{\psi_S}$ is prepared at a rate $\kappa_{{\rm eff},S} \equiv |\bra{\psi_S} \hat{L}^\kappa_{\rm eff} \ket{T}|^2=\frac{4 b^2 g^2 \Omega^2}{\Delta^2 \kappa \left(2 b^2+\Omega_{\rm MW}^2\right)}$, ($\kappa_{{\rm eff},S} \ll \kappa_{\rm eff}$) and decays only by spontaneous emission.\\
For the derivation of the error scaling we use a rate argument to compare the decay rates into and out of the steady state ($\dot{P}=0$, $P_{\psi_S} \approx 1$)
\begin{eqnarray}
\left(1-F_{\psi_S}\right) \approx 3 P_T \approx \frac{3 \Gamma_{\psi_S \rightarrow}}{\Gamma_{\rightarrow \psi_S}} P_S \approx \frac{3 \gamma \kappa \left(4 b^2+3 \Omega_{\rm MW}^2\right) \Omega_{\rm MW}^2}{64 g^2 \left(2 b^2+\Omega_{\rm MW}^2\right) b^2},
\end{eqnarray}
where we have used the strong coupling condition, and the assumption that the populations of the three undesired states are well-shuffled by $\Omega_{\rm MW}$ so that they have a similar population. In contrast to the previously presented schemes, the static error of the protocol incorporates a second term that determines the preset compromise in the fidelity of the steady state due to the asymmetry $b$ so that 
\begin{eqnarray}
\left(1-F_{S}\right) &= \left(1-F_{\psi_S}\right) + \left(1-\left| \langle \psi_S | S \rangle \right|^2\right) = \nonumber \\ &= \frac{3 \gamma  \kappa \left(4 b^2+3 \Omega_{\rm MW}^2\right) \Omega_{\rm MW}^2}{64 g^2 \left(2 b^2+\Omega_{\rm MW}^2\right) b^2} + \frac{2 b^2}{2 b^2+\Omega_{\rm MW}^2}.
\end{eqnarray}
The minimal overall error is reached for a trade-off at which these terms are equal. This compromise between establishing the steady state by the asymmetry, and at the same time avoiding the decrease in its fidelity by the asymmetry, is the cause of the different scaling of the error and speed discussed in Sec. \ref{SectionComparison}. For the parameter $b$ we obtain the condition
\begin{eqnarray}
b_{\rm opt} = \frac{\sqrt{3}\Omega_{\rm MW}}{\sqrt[4]{2^5}} \sqrt[4]{\frac{\gamma \kappa}{g^2}}.
\end{eqnarray}
Inserting this yields the effective decay rate into $\ket{\psi_S}$
\begin{equation}
\kappa_{{\rm eff},S}=\frac{4 g^2 \Omega^2}{\Delta^2 \kappa \left(2 + \frac{3}{\sqrt{32 C}}\right)} \approx \frac{2 g^2 \Omega^2}{\Delta^2 \kappa}
\end{equation}
and the error of the protocol
\begin{eqnarray}
\left(1-F_S\right)_{\rm WS}=\frac{3 \gamma  \kappa  \left(8 g+\sqrt{2} \sqrt{\gamma  \kappa }\right)}{4 g \left(3 \gamma  \kappa +4 \sqrt{2} g \sqrt{\gamma  \kappa }\right)} \approx \frac{3}{2 \sqrt{2C}}.
\end{eqnarray}
Thus, we find that for the WS protocol the fidelity of the steady state with the maximally entangled singlet state exhibits a scaling with one over the square root of the cooperativity of the cavity. As with the $\ket{S_1}$ scheme, we have one prominent decay process to prepare the singlet from the three dressed ground states. With $b_{\rm opt}$ the spectral gap is then given by
\begin{equation}
\lambda_{\rm WS} = \frac{\kappa_{{\rm eff},S}}{3} = \frac{2 g^2 \Omega^2}{3 \Delta^2 \kappa}.
\end{equation}
A numerical comparison with the other schemes is given in Sec. \ref{SectionComparison}.

\begin{figure*}[ht]
\centering
\includegraphics[width=8cm]{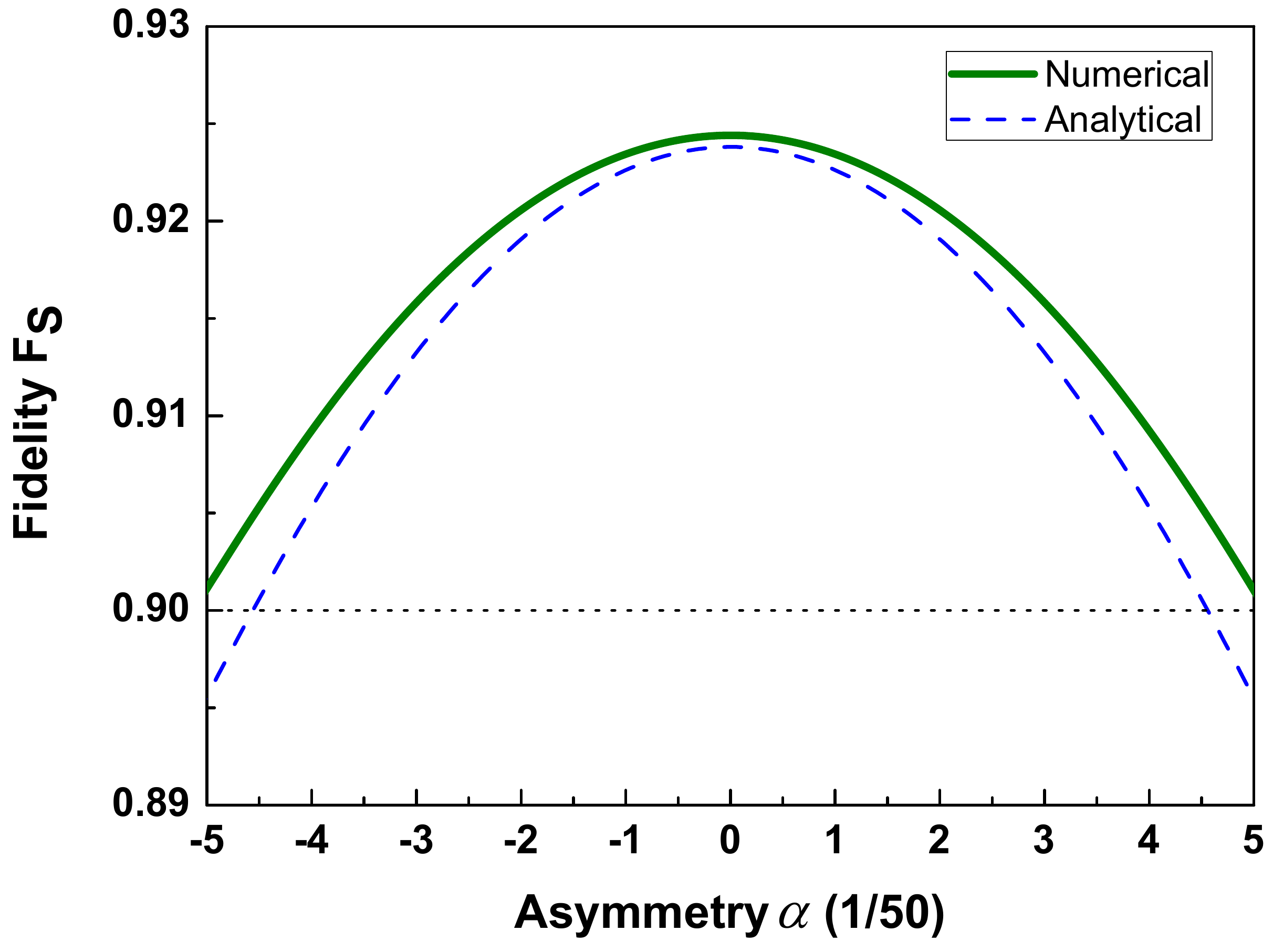}
\caption{Fidelity under asymmetric coupling to the cavity. Numerical results from the full Liouvillian (solid green) are well-approximated by the analytic findings (blue dash). High fidelities of about $90 \%$ are achieved up to $|\alpha| \approx 0.1$, where $\frac{g_1}{g_2}\approx1.22$.}
\label{FigAsymmetry}
\end{figure*}

\section{Appendix B: Effects from imperfect coupling of the atoms to the cavity}
\label{AppAsymm}
Experimental realization of a scheme for dissipative state preparation requires an understanding of the effects originating from the imperfect couplings of the atoms to the cavity mode. In state-of-the-art cavity QED systems, such as Ref. \cite{Kubanek1}, longitudinal confinement prevents fluctuation of the atomic positions along the cavity axis. Still, a static difference in the couplings of the two atoms is possible. Expressing these couplings as $g_1=g(1+\alpha)$ and $g_2=g(1-\alpha)$ the above analyses can still be carried out for the mean coupling of $g = \frac{1}{2} \left(g_1 + g_2\right)$. From the asymmetry $\alpha$ an additional source of error emerges. Below, we exemplarily derive this asymmetry error for the $\ket{S_1}$ scheme. In case of a static, asymmetric coupling of the two atoms to the cavity the atom-cavity coupling can be written as
\begin{eqnarray}
\hat{H}_{\rm ac}&=\hat{a} \left(g_1 \ket{e}_1\bra{1} + g_2 \ket{e}_2\bra{1}\right) + H.c. \nonumber \\
&=\hat{a} g \left(\left(1+\alpha\right)  \ket{e}_1\bra{1} + (1-\alpha) \ket{e}_2\bra{1}\right) + H.c.,
\end{eqnarray}
The asymmetry error affects both the dynamics of the populations and the coherences so that we use the effective Liouvillian $\mathcal{L}_{\rm eff}$ to derive the steady state, after having excluded other sources of error beforehand ($\kappa_{\rm eff} \rightarrow 0$). For weak driving and strong coupling $g \gg \left(\gamma, \kappa\right) \gg \left(\Omega, \Omega_{\rm MW}, \beta\right)$ the asymmetry error $\alpha$ can be effectively decoupled from both the static error ($\propto C^{-1}$) and the dynamic error ($\propto \Omega^{2}$) and we obtain
\begin{eqnarray}
\left(1-F_S\right)_\alpha \approx 3 \alpha ^2.
\end{eqnarray}
The result is plotted in Fig. \ref{FigAsymmetry} using the parameters of Ref. \cite{Kubanek1}. The effect of an asymmetric coupling is found to be rather small as compared to other sources of errors. For $|\alpha|\approx 0.1$ the loss of fidelity through asymmetry is as little as $\approx 2 \%$. In this case, with $\frac{g_1}{g_2}\approx1.22$, fidelities of about $90 \%$ are still achievable.

\begin{figure*}[ht]
\centering
\includegraphics[width=15cm]{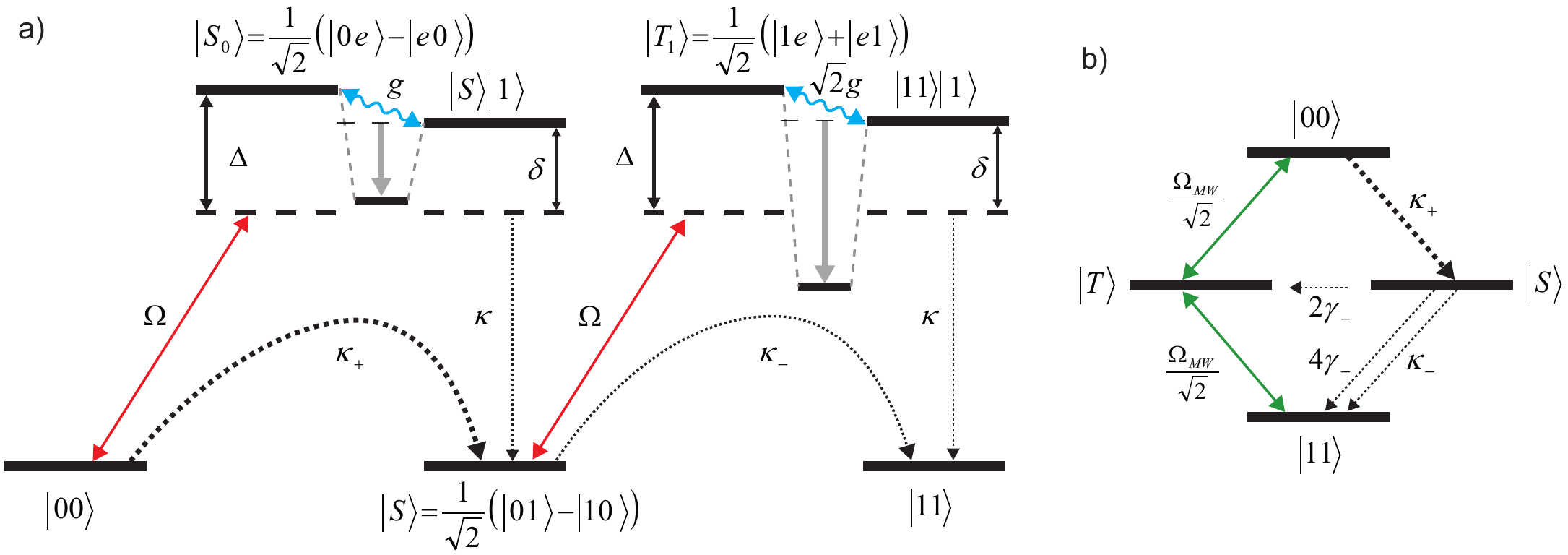}
\caption{Mechanism and effective processes of the $\ket{S_0}$ scheme. (a) Setting $\delta=\frac{g^2}{\Delta}$ shifts the lower dressed state of $\ket{S_0}$ and $\ket{S}\ket{1}$ into resonance with the optical driving $\Omega$, while the dressed states of $\ket{T_1}$ and $\ket{11}\ket{1}$ are detuned due to a different coupling strength of $\sqrt{2} g$. Hence, effective cavity decay $\kappa_+$ from $\ket{00}$ into $\ket{S}$ is enhanced while cavity loss $\kappa_-$ from $\ket{S}$ into $\ket{11}$ is suppressed. The populations of the triplet states are shuffled by a microwave or Raman transition $\Omega_{\rm MW}$, shown together with the effective decay processes in (b).}
\label{FigS0Mechanism}
\end{figure*}

\section[Appendix C: A scheme for directed cavity decay]{Appendix C: A scheme for directed cavity decay via $\ket{S_0}$}
\label{AppS0}
In Ref. \cite{KRS} we presented a scheme for the dissipative preparation of entanglement that employs strongly engineered cavity decay to prepare the singlet state $\ket{S}$. In the following we corroborate the claims made about error and speed in Ref. \cite{KRS}. For this $\ket{S_0}$ scheme, we will drive the atoms with opposite phase $\phi=\pi$ so that we can conduct our discussion based on the effective operators of Eqs. (\ref{EqS1EO}-\ref{EqS1EO3}) previously derived for the $\ket{S_1}$ scheme.\\
In brief, the mechanism, as visualized in Fig. \ref{FigS0Mechanism}, is the following: Population from state $\ket{00}$ is driven up to the excited state $\ket{S_0}=\frac{1}{2}\left(\ket{0e}-\ket{e0}\right)$, at a laser detuning of $\Delta$ with $\beta=0$. State $\ket{S_0}$ is in turn coupled by the atom-cavity interaction to $\ket{S}\ket{1}$ with a strength of $g$. $\ket{S}\ket{1}$ decays into $\ket{S}$ via cavity decay at a rate of $\kappa$. Setting the cavity detuning to $\delta=\frac{g^2}{\Delta}$ greatly enhances the effective cavity decay $\ket{00} \overset{\Omega}{\rightarrow} \ket{S_0} \overset{g}{\rightarrow} \ket{S}\ket{1} \overset{\kappa}{\rightarrow} \ket{S}$. As in the $\ket{T_{0,1}}$ schemes, this is due to the fact that the lower dressed state of $\ket{S_0}$ and $\ket{S}\ket{1}$ is shifted into resonance. Loss of population from the singlet via $\ket{T_1}$ is once again effectively suppressed, since $\ket{T_1}$ and $\ket{11}\ket{1}$ are coupled with a larger strength $\sqrt{2} g$, shifting the dressed states out of resonance. The triplet states are shuffled by a microwave or Raman field with optimal strength $\Omega_{\rm MW}\approx \Omega/3$.\\
We find that the optimal atomic detuning is  $\Delta = g \sqrt{\frac{\gamma}{\kappa}}$. At this detuning we obtain the effective operators
\begin{eqnarray}
\hat{L}_{\rm eff}^{\kappa} &= \sqrt{\kappa_-}\ket{11}\bra{S}+i\sqrt{\kappa_+} \ket{S}\bra{00} \\
\hat{L}_{\rm eff}^{\gamma,0,\{1,2\}} &= \pm i \sqrt{2 \gamma_+} \ket{00}\bra{00}-\sqrt{\gamma_-} \left(\ket{T}\bra{S} \pm \ket{S}\bra{S} \right)\\
\hat{L}_{\rm eff}^{\gamma,1,\{1,2\}} &= - \sqrt{2\gamma_-} \ket{11}\bra{S} + i \sqrt{\gamma_+} \left( \pm \ket{T}\bra{00} - \ket{S}\bra{00}\right),
\end{eqnarray}
where we have assigned $\kappa_+= \frac{\Omega^2}{2 \gamma}= 8 \gamma_+$, $\kappa_-=\frac{\kappa \Omega^2}{2 g^2}=16 \gamma_-$.
Indeed, the most pronounced process is the strongly enhanced effective cavity decay from $\ket{00}$ into $\ket{S}$.
The static error and the spectral gap are found to be
\begin{eqnarray}
\left(1-F_S\right)_{\ket{S_0}} &= \frac{7}{2 C} \\
\lambda_{\ket{S_0}} &=\frac{5 -\sqrt{5}}{16}\frac{\Omega^2}{\gamma}.
\end{eqnarray}
A comparative numerical study of the performance of this scheme is given in Sec. \ref{SectionComparison}.

\end{document}